\documentclass[twocolumn]{aastex631}  % , linenumbers

\shorttitle{Gravitational lensing formalism in a curved arc basis}
\shortauthors{S. Birrer}
\graphicspath{{./}{figures/}}
\usepackage{amsmath}    % Advanced maths commands
\usepackage{bbold}  % for identiy matrix
\usepackage{fontawesome}  % for GitHub symbole

\begin{document}

\title{Gravitational lensing formalism in a curved arc basis: A continuous description of observables and degeneracies from the weak to the strong lensing regime}

\author[0000-0003-3195-5507]{Simon Birrer}
\altaffiliation{sibirrer@stanford.edu}
\affiliation{Kavli Institute for Particle Astrophysics and Cosmology and Department of Physics,\\
 Stanford University, Stanford, CA 94305, USA}
\affiliation{SLAC National Accelerator Laboratory, Menlo Park, CA, 94025, USA}



\begin{abstract}

% statement of need
Gravitationally lensed curved arcs provide a wealth of information about the underlying lensing distortions. Extracting precise lensing information from extended sources is a key component in many studies aiming to answer fundamental questions about the Universe. To maintain accuracy with increased precision, it is of vital importance to characterize and understand the impact of degeneracies inherent in lensing observables.
In this work, we present a formalism to describe the gravitational lensing distortion effects resulting in curved extended arcs based on the eigenvectors and eigenvalues of the local lensing Jacobian and their directional differentials.
We identify a non-local and non-linear extended deflector basis that inherits these local properties.
Our parameterization is tightly linked to observable features in extended sources and allows one to accurately extract the lensing information of extended images without imposing an explicit global deflector model.
We quantify what degeneracies can be broken based on specific assumptions on the local lensing nature and assumed intrinsic source shape.
Our formalism is applicable from the weak linear regime, the semi-linear regime all the way up to the highly non-linear regime of highly magnified arcs of multiple images.
The methodology and implementation presented in this work provides a framework to assessing systematics, to guide inference efforts in the right choices in complexity based on the data at hand, and to quantify the lensing information extracted in a model-independent way. \href{https://github.com/sibirrer/curved_arcs}{\faGithub}

\end{abstract}

\keywords{Gravitational lensing (670) --- stong gravitational lensing (1643) --- weak gravitational lensing (1797)}


\section{Introduction} \label{sec:introduction}
Gravitational lensing displaces the observed position and distorts the shape of apparent objects on the sky due to intervening inhomogeneous matter along the line of sight.
In the cosmological context, the lensing effect can mostly be well approximated with a first order displacements and second order perturbations on the shape of the lensed source \citep[see e.g.,][]{Blandford:1992, Kaiser:1993, Kaiser:1995, Mellier:1999, Bartelmann:2001}. The displacement effect is not an observable as the intrinsic position of the objects can not be determined. The distortion of the shapes of extended objects do contain statistical signal due to the correlation of the apparent shapes of different objects along similar lines of sights, known as cosmic shear.

In the very close vicinity of massive over-densities, such as galaxies or galaxy clusters, the lensing effect can lead to highly distorted images and even the appearance of multiple images of the same source. In these regimes, the second order perturbations do not accurately describe the observed distortions of extended lensed sources anymore.

One way to bridge the gap between the linear and non-linear lensing distortion regimes is with third order polynomial perturbations on the lensing potential (flexion) as an octopole signal in the measured shape \citep[e.g.][]{Goldberg:2002, Goldberg:2005, Irwin:2005, Irwin:2006, Bacon:2006}.
The flexion measurement has been employed, for example, in the Hubble Ultra Deep Field \citep{Irwin:2007}, and in combination with shear and strong lensing conjugate points in cluster models, both in parametric and non-parametric form \citep[e.g.,][]{Leonard:2007}. 
A Taylor expansion determination of lensing quantities to fourth order has been investigated by \cite{WagnerBartelmann:2016}, and a generalized weak lensing effect by \cite{Fleury:2019}. Overall, there has been only moderate success and applicability of flexion corrections in providing model-independent local lensing constraints.

In certain regimes, a fourth order approximation with a carefully chosen coordinate system can match some further positional and local constraints in quadruply imaged lenses \citep{Wagner:2019Universe} but still does not allow one to describe extended arcs accurately.
The reason for the limits in polynomial extensions is the non-perturbative nature beyond the second order (shear and convergence) of the matter distribution in the Universe per se, leading to non-linear lensing effects deviating from a Taylor expansion \citep[e.g.,][]{Schneider:2008}. The most prominent and abundant signature of non-linear lensing effects beyond shear and convergence are curved arc distortions. 
Thought, in the infinitesimal small sources, the alignment of quadrupole and octupole moments induce curvature locally \citep[e.g.,][]{Irwin:2005}, more extended observed effects can not be described by third order flexion terms (or even higher order ones) and require a non-perturbative treatment of the lensing effect.

In the absence of a clean data-driven approach in the non-linear regime, the use of explicit deflector mass models to provide the link between the observables and the lensing deflection field became the standard in many analyzes involving strong gravitational lensing.
One example of a lens model family widely employed is the singular power-law mass density profile. For the spherical case, there are theoretical studies quantifying and discussing how well observables are able to constrain the global mass profile slope of the imposed power-law profile \citep[e.g.][]{Suyu:2012pl, ORiordan:2019}.

The constraints derived from parameterized models may not in all circumstances reflect the observational information on the deflection field. When employing a specific parameterized model, constraints can be derived within the specific lens model family parameters only.
In the case of the constant power-law slope mass profile, the constraints on the logarithmic slope from extended imaging observables are are only possible due to the demanding constraints of the global deflector model. Neither the local slope nor the average of the sloe within a certain range are observables themselves per se.

On one hand, specific functional forms may only probe a sub-set of possible lensing configuration allowed by the data, leading to over-constrained deflector inferences. For instance, an imposed functional form on the deflector profile can artificially break the mass-sheet degeneracy \citep[MST,][]{Falco:1985no, Gorenstein:1988} and potentially bias the inference of the Hubble constant, $H_0$, from time-delay cosmography measurements \citep[e.g.][]{Schneider:2013pm, Birrer:2016zy, Sonnenfeld:2018, Kochanek:2020a, Birrer:2020}.
We refer to \cite{Birrer:2020} for the latest constraints on $H_0$ by the TDCOSMO collaboration when only using MST-invariant imaging observables by effectively allowing an additional MST degree of freedom in the mass profiles and anchoring the radial density profiles by stellar kinematics measurements.

On the other hand, a certain model may be insufficient in describing the wealth of data available. This can particularly be the case in the galaxy cluster regime where parameterized models are currently limited to match conjugate points within the astrometric measurement uncertainty and are incapable to describe the relative distortions observed in extended sources to the noise level of high resolution data \citep{Yang:2020, Dai:2020}. 
Another example is the interpretation of anomalous quadruply lensed quasar flux ratios. In some cases, observed flux ratios may not exclusively be due to dark matter substructure but instead might have contributions from from larger scale baryonic components in the lensing galaxy that a simplified lens model may have neglected \citep[see e.g.,][]{Hsueh:2016, Hsueh:2018, Gilman:2017}. Similar effects can be observed when quantifying distortions in extended arcs, e.g. in \cite{Birrer:2017} larger scale potential corrections had to be applied before substructure investigations could proceed.

The key to extract maximal precision while maintaining accuracy in the non-linear regime of gravitational lensing is to allow for freedom in the lensing description where data is able to constrain it and to have transparent priors in regimes where the data does not provide information to the problem at hand.
The aim of this paper is to provide a theoretical formalism that allows one to quantify the invariant observables in gravitational lensing and a practical implementation to extract this information from extended lensed images.

We introduce a formalism to describe the distortion effects of curved extended arcs based on the eigenvectors and eigenvalues of the local lensing Jacobian and their directional differentials. The eigenvectors and their differentials are describing particular aspects of observational lensing features.
We identify specific bases for a non-local non-linear extension of the local properties to accurately predict and describe the detailed shape of extended sources at and around the location of interest without the need of a globally defined deflector model.

Degeneracies inherent in lensing, such as the MST and it's generalization, the Source Position Transform (SPT) \citep{Schneider:2014, Unruh:2017, Wertz:2018} pose limits on the extractable lensing information. In the most general form, the SPT is not restricted to curl-free deflector fields. The arc basis introduced in this work is a suited approach to explore the curl-free components of the SPT with some minimal, but well motivated, broad assumptions on the local lensing distortions.
The method presented in this work effectively allows one to extract lensing informations from extended sources mitigating degeneracies inherent in lensing.

Our formalism is applicable in all cosmological regimes of gravitational lensing, from the weak linear regime, the semi-linear regime, up to the fully non-linear regime of highly magnified arcs and Einstein rings of multiple images.

The paper is organized as follows: In Section~\ref{sec:arc_distortions} we introduce the formalism of radial and tangential distortions in the eigenvector basis of the lensing Jacobian and their differentials.
We then discuss local lensing invariances and degeneracies in the context of the curved arc basis in Section~\ref{sec:invariances}.
In Section~\ref{sec:profile_constraints} we discuss the observables in curved arcs that allow us to constrain radial and tangential aspects of a global mass distributions and demonstrate how our formalism is able to extract all relevant information without being over-constraint.
In Section~\ref{sec:application} we elaborate about applications of the methodology that can benefit from the approach we introduce in this work, its limitations, and provide a specific example.
We conclude in \ref{sec:conclusion}.

All figures and inferences can be reproduced using code available at \href{https://github.com/sibirrer/curved_arcs}{this repository~\faGithub}\footnote{\url{https://github.com/sibirrer/curved_arcs}}. All numerical computations are performed with \textsc{lenstronomy}\footnote{\url{https://github.com/sibirrer/lenstronomy}} \citep{Birrer_lenstronomy, lenstronomyII} version \textsc{1.8.2}.

\section{Lensing formalism for curved arcs} \label{sec:arc_distortions}
In this section, we first review the lensing formalism in general terms, in particular the polynomial Cartesian expansion in second and third order differentials of the lensing potential (\ref{sec:lensing_basics}). We then introduce the formalism of the differentials in the eigenvector basis (\ref{sec:eigenvector}). We identify the local differentials in eigenvector space attributed to curved arcs which provide a continuous mapping from the weak lensing to the strong lensing regime (\ref{sec:tangential_arcs}).
We use the eigenvector basis to define a minimal local lens model able to describe extended curved arcs preserving the key differential quantities over an extended area around the localized position (\ref{sec:curved_arc_basis}).

\subsection{Lensing formalism basics} \label{sec:lensing_basics}

\subsubsection{Lens equation}
The lens equation, which describes the mapping from the source plane $\boldsymbol{\beta}$ to the image plane $\boldsymbol{\theta}$, is given by
\begin{equation} \label{eqn:lens_equation}
  \boldsymbol{\beta} = \boldsymbol{\theta} - \boldsymbol{\alpha}(\boldsymbol{\theta}),
\end{equation}
where $\boldsymbol{\alpha}$ is the angular deflection as seen on the sky between the original unlensed and the lensed observed position of an object.

\subsubsection{First and second order Cartesian differentials}
The differential of the lens equation between the source position and its lensed appearance, the Jacobian, is
\begin{equation} \label{eqn:jacobian}
A_{ij} \equiv \frac{\partial \beta_i}{\partial \theta_j}=\delta_{ij} - \frac{\partial \alpha_i}{\partial \theta_j}.
\end{equation}
The Jacobian describes the local linear distortions of a small extended source or likewise the magnification of an unresolved small source. The magnification $\mu$ is the change in differential area from the source to the image position and can be expressed as the determinant of the inverse Jacobian
\begin{equation} \label{eqn:magnification_general}
	\mu = \text{det}(\boldsymbol{A})^{-1}.
\end{equation}
The components of the Jacobian can be decomposed into the convergence
\begin{equation}
	\kappa = \frac{1}{2} \left( \frac{\partial \alpha_x}{\partial \theta_x} + \frac{\partial \alpha_y}{\partial \theta_y}\right),
\end{equation}
the shear components
\begin{equation}\label{eqn:gamma1}
	\gamma_1 = \frac{1}{2} \left( \frac{\partial \alpha_x}{\partial \theta_x} - \frac{\partial \alpha_y}{\partial \theta_y}\right),
\end{equation}

\begin{equation}\label{eqn:gamma2}
	\gamma_2 = \frac{1}{2} \left( \frac{\partial \alpha_x}{\partial \theta_y} + \frac{\partial \alpha_y}{\partial \theta_x}\right),
\end{equation}
and the curl component
\begin{equation}
	{\rm curl} = \left( \frac{\partial \alpha_x}{\partial \theta_y} - \frac{\partial \alpha_y}{\partial \theta_x}\right).
\end{equation}

The next to leading order polynomial expansion of the lens equation is known as flexion \citep{Goldberg:2002, Goldberg:2005, Bacon:2006} and describes the gradients of the Jacobian
\begin{equation} \label{eqn:flexion_differentials}
	D_{ijk} \equiv \frac{\partial A_{ij}}{\partial \theta_k}.
\end{equation}
The lens equation (Eqn \ref{eqn:lens_equation}) to second polynomial order in $\boldsymbol{\theta} = \boldsymbol{\theta}_0 + \boldsymbol{\Delta\theta}$ is given by
\begin{equation}
	\beta_i \approx \theta_i - \alpha_i\left(\boldsymbol{\theta}_0\right) + A_{ij} \Delta\theta_j + \frac{1}{2}D_{ijk} \Delta\theta_j\Delta\theta_k.
\end{equation}
To this stage, no symmetry on the form of the lens equation (Eqn \ref{eqn:lens_equation}) or the Jacobian (Eqn \ref{eqn:jacobian}) have been invoked.

In the case of a single lensing plane, the source term of the gravitational deflection field is the convergence field, $\kappa(\boldsymbol{\theta })$, with zero curl, and there exists a scalar lensing potential, $\psi$, given by
\begin{equation}\label{eqn:potential}
\psi ({\boldsymbol{\theta }})={\frac {1}{\pi }}\int d^{2}\boldsymbol{\theta ^{\prime }}\kappa ({\boldsymbol{\theta }}^{\prime })\ln |{\boldsymbol{\theta }}-{\boldsymbol{\theta }}^{\prime }|,
\end{equation}
such that
\begin{equation}
    \boldsymbol{\alpha}(\boldsymbol{\theta}) = \boldsymbol{\nabla} \psi(\boldsymbol{\theta}).
\end{equation}
The Jacobian (Eqn \ref{eqn:jacobian}) is symmetric, without any curl component, and can be decomposed into a trace (convergence $\kappa$) and trace-free (shear $\gamma_1$, $\gamma_2$) term as
\begin{equation}
A_{ij} = \delta_{ij} - \frac{\partial^2 \psi}{\partial \theta_i \partial \theta_j} \equiv \left[\begin{array}{ c c } 1-\kappa -\gamma_1 & -\gamma_2 \\ -\gamma_2 & 1-\kappa +\gamma_1 \end{array}\right].
\end{equation}
The magnification $\mu$ (Eqn \ref{eqn:magnification_general}) can be written as
\begin{equation}
	\mu = \frac{1}{(1-\kappa)^2 - \gamma_1^2 - \gamma_2^2}.
\end{equation}

The flexion terms can be compactly written as \citep[see e.g.][]{Kaiser:1998}
\begin{equation}
D_{ij1} =  \left[\begin{array}{ c c } -2\gamma_{1, 1} - \gamma_{2, 2} & -\gamma_{2, 1} \\ -\gamma_{2, 1} & -\gamma_{2, 2} \end{array}\right].
\end{equation}
\begin{equation}
D_{ij2} =  \left[\begin{array}{ c c } -\gamma_{2, 1} & -\gamma_{2, 2} \\ -\gamma_{2, 2} & 2\gamma_{1, 2} -\gamma_{2, 1} \end{array}\right].
\end{equation}

\subsection{Differentials in eigenvector space notation} \label{sec:eigenvector}

\subsubsection{Jacobian in eigenvector space} \label{sec:eigenvector_jacobian}
We describe the Jacobian $\mathbf{A}$ (Eqn.~\ref{eqn:jacobian}) in terms of its two eigenvectors $\boldsymbol{e}_i$ with corresponding eigenvalues $\lambda_i$

\begin{equation}
	\mathbf{A} \cdot \boldsymbol{e_i} = \lambda_i \boldsymbol{e}_i.
\end{equation}
In the case of a symmetric Jacobian, the eigenvectors are orthogonal and the eigenvalues are real.
The magnification $\mu$ (Eqn.~\ref{eqn:magnification_general}) can be written as
\begin{equation} \label{eqn:magnification_eigenvector_general}
	\mu = \prod_{i}1/\lambda_i.
\end{equation}
In the weak lensing regime, the two eigenvectors and the direction provide a complete and equivalent description to shear and convergence and we can state the properties by referring to the major and minor eigenvector.

In the vicinity of a collapsed over-dense structure, such as a galaxy or a galaxy cluster, the two eigenvectors are to good approximation radial and tangential to the center of the structure. We can associate the eigenvalues as the inverse radial and tangential stretch of an image exhibited by the massive structure
\begin{align*} 
	\mathbf{A} \cdot \boldsymbol{e_{\text{rad}}} &= \lambda_{\text{rad}}^{-1} \boldsymbol{e}_{\text{rad}} \\
	\mathbf{A} \cdot \boldsymbol{e_{\text{tan}}} &= \lambda_{\text{tan}}^{-1} \boldsymbol{e}_{\text{tan}},
\end{align*} 
where we noted $\boldsymbol{e}_{\text{rad}}$ to be the radial component and $\boldsymbol{e}_{\text{tan}}$ to be the tangential component of the Jacobian $\mathbf{A}$ with their corresponding eigenvalues $\lambda_{\text{rad}}^{-1}$ and $\lambda_{\text{tan}}^{-1}$.
In this form, $\lambda_{\text{rad}}$ corresponds to the stretch factor of the source in radial direction and $\lambda_{\text{tan}}$ in tangential direction, corresponding to
\begin{align}
	\lambda_{\rm rad} = \frac{\partial \theta_r}{\partial \beta_r}, &&
	\lambda_{\rm tan} = \frac{\partial \theta_t}{\partial \beta_t}, 
\end{align}
where $\partial\beta_r$ ($\partial\beta_t$) correspond to the directional differentials in the source plane corresponding to the reflected radial (tangential) direction in the image plane.
The magnification is the product of the orthogonal stretches
\begin{equation} \label{eqn:magnification_eigenvector}
	\mu = \lambda_{\text{rad}} \lambda_{\text{tan}}.
\end{equation}

We define the scalar angle $\phi_{\text{tan}}$ ($\phi_{\text{rad}}$) as the angle between the eigenvector $\boldsymbol{e_{\text{tan}}}$ ($\boldsymbol{e_{\text{rad}}}$) and a specific polar coordinate system of choice (e.g. centered at the massive structure for convenience) such that
\begin{equation}
	\text{cos}\left(\phi_{\text{tan}}\right) = \boldsymbol{e}_{\text{tan}} \cdot \boldsymbol{e}_{\text{0}}
\end{equation}
with $\boldsymbol{e}_{\text{0}}$ is the unit vector in the direction of the coordinate center. A convenient coordinate center is the center of a mass distribution.

In general terms, we can associate the tangential direction to be along the major shear direction and the radial component orthogonal to it. The directions of the eigenvectors themselves are independent of the coordinate center.

\subsubsection{Third order differentials in eigenvector space} \label{sec:eigenvector_differentials}
Analogously to the polynomial flexion as the differentials of the Jacobian in Cartesian direction, we can introduce differentials along the eigenvectors of the tangential and radial eigenvalues as well as the differentials of the direction of the eigenvectors themselves.

The differential of the eigenvector directions along its own direction provides a measure of curvature. We define the tangential curvature, $s_{\text{tan}}$, as
\begin{equation} \label{eqn:stang}
	s_{\text{tan}} \equiv \frac{\partial \phi_{\text{tan}}}{ \partial \boldsymbol{e_{\text{tan}}}}
\end{equation}
and the curvature in the radial direction, $s_{\text{rad}}$, as
\begin{equation} \label{eqn:srad}
	s_{\text{rad}} \equiv \frac{\partial \phi_{\text{rad}}}{ \partial \boldsymbol{e_{\text{rad}}}}.
\end{equation}

The curvature terms $s_{\text{tan}}$ and $s_{\text{rad}}$ are coordinate system independent. We note that the directional differentials in the two directions $\boldsymbol{e_{\text{tan}}}$ and $\boldsymbol{e_{\text{rad}}}$ are the same when the eigenvectors are orthogonal.

For the differentials of the eigenvalues we introduce the following notations. The gradient of the tangential magnification in tangential direction is
\begin{equation}
	\partial_t\lambda_{\rm tan} \equiv \frac{\partial \lambda_{\rm tan}}{\partial \boldsymbol{e}_{\rm tan}},
\end{equation}
and in radial direction
\begin{equation}
	\partial_r\lambda_{\rm tan} \equiv \frac{\partial \lambda_{\rm tan}}{\partial \boldsymbol{e}_{\text{rad}}}.
\end{equation}
The gradient of the radial magnification in radial direction is
\begin{equation}
	\partial_r\lambda_{\rm rad} \equiv \frac{\partial \lambda_{\rm rad}}{\partial \boldsymbol{e}_{\text{rad}}}
\end{equation}
and in tangential direction is
\begin{equation} \label{eqn:s_tan}
	\partial_t\lambda_{\rm rad} \equiv \frac{\partial \lambda_{\text{rad}}}{\partial \boldsymbol{e}_{\text{tan}}}.
\end{equation}

\subsection{Tangential arcs and their eigenvector components} \label{sec:tangential_arcs}
\begin{figure}
  \centering
  \includegraphics[angle=0, width=80mm]{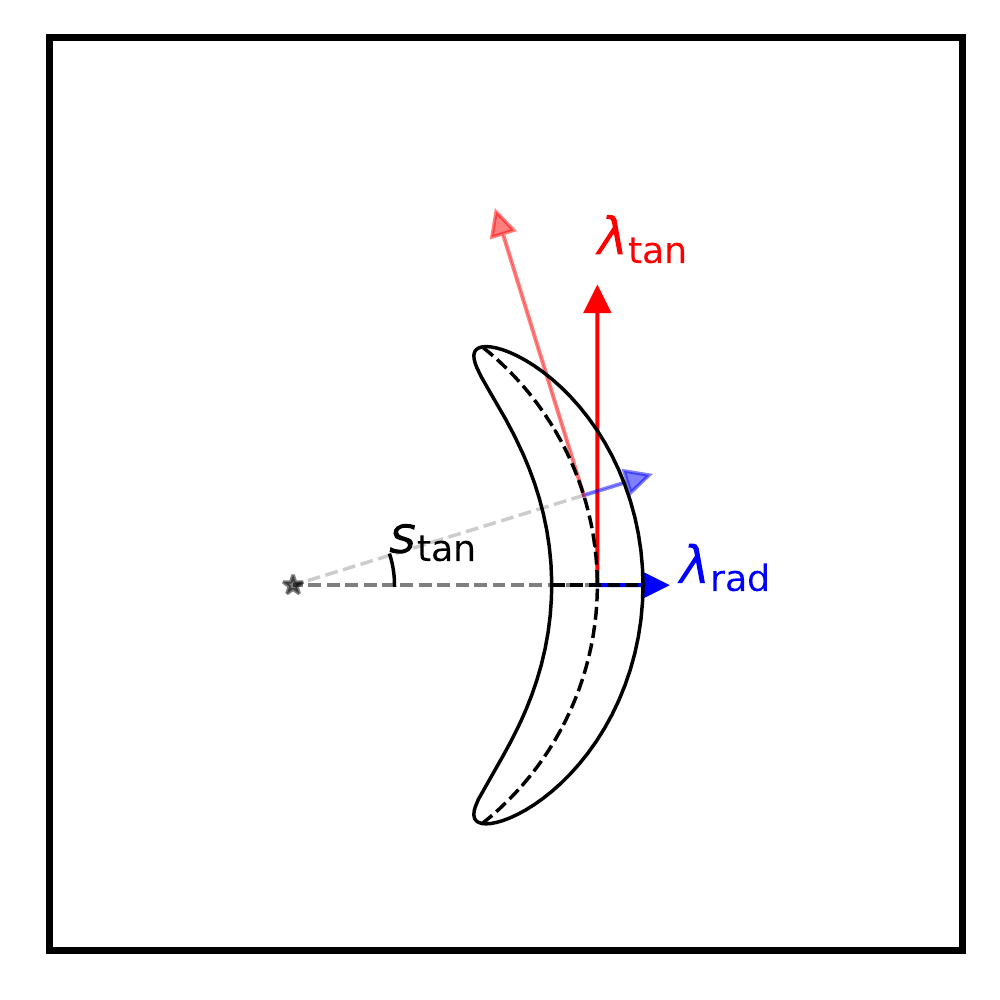}
  \caption{Illustration of eigenvectors and curvature along the tangential direction for a single arc feature. The directional differential along the tangential direction $s_{\rm tan}$ marks a radius with radius $r = s_{\rm tan}^{-1}$ on which the tangential and radial eigenvalues, $\lambda_{\rm tan}$ and $\lambda_{\rm rad}$ are constant and pointing either in the direction or orthogonal to the center. \href{https://github.com/sibirrer/curved_arcs/blob/v1.0/Notebooks/curved_arc_illustration.ipynb}{\faGithub}}
\label{fig:eigenvector_illustration}
\end{figure}

In the following we focus on a single, considered most prominent, third order eigenvector differential, the curvature in the direction of the tangential direction $s_{\rm tan}$ (Eqn.~\ref{eqn:s_tan}).
We can describe in a minimal form a tangential arc by considering radial and tangential stretch, $\lambda_{\text{rad}}$ and $\lambda_{\text{tan}}$, radial direction, $\phi_{\rm rad}$, and the curvature in the tangential direction, $s_{\rm tan}$ (Eqn.~\ref{eqn:s_tan}).
Figure~\ref{fig:eigenvector_illustration} provides an example of an arc fully described by these four components. The eigenvalues remain constant along a circle defined by the inverse curvature $r = s_{\rm tan}^{-1}$.

\begin{figure*}
  \centering
  \includegraphics[angle=0, width=130mm]{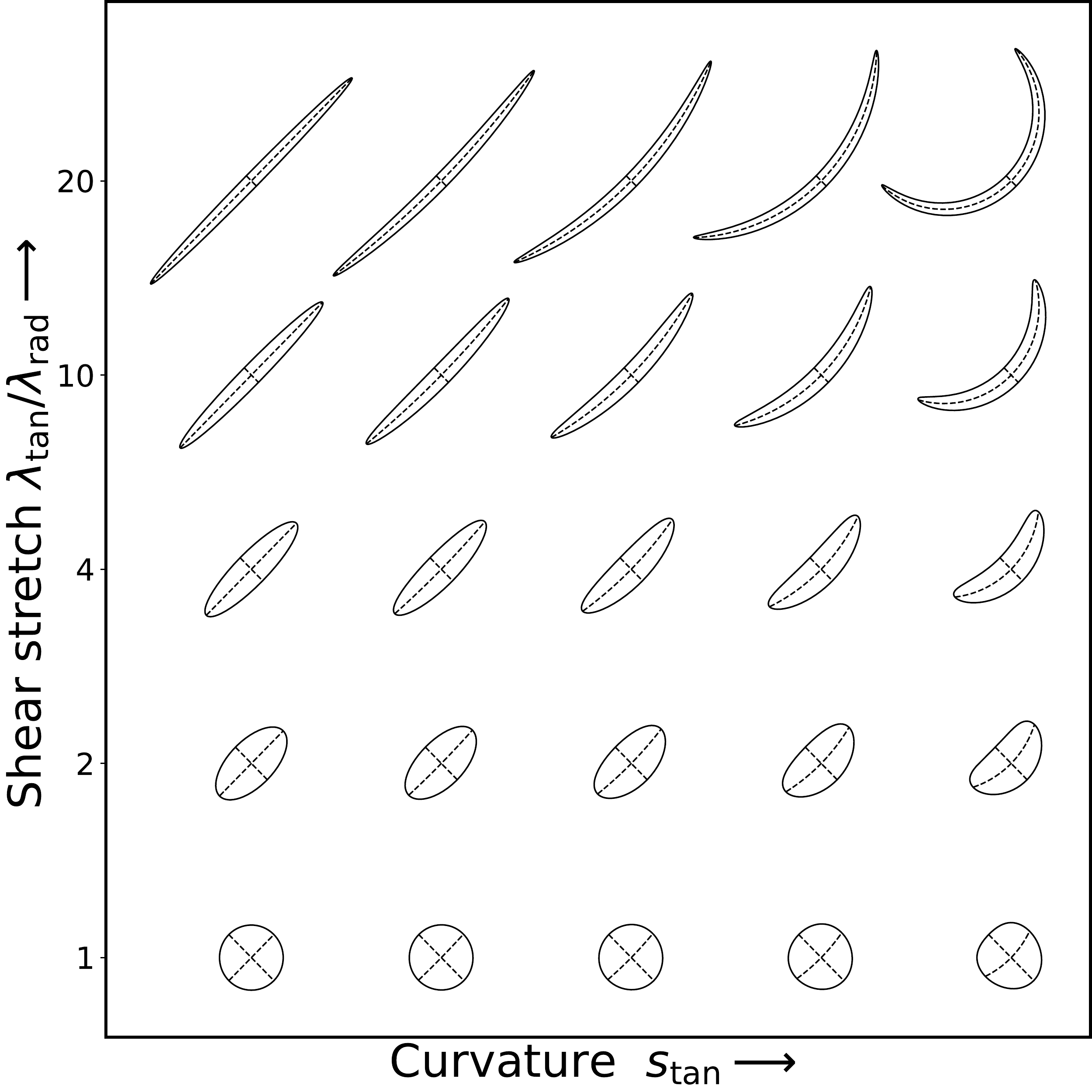}
  \caption{Illustration of tangential arcs as a function of tangential to radial eigenvector stretch ratio $\lambda_{\rm tan} / \lambda_{\rm rad}$ and tangential curvature $s_{\rm tan}$. The description of curved arcs in the eigenvector components allow us to describe distortions of lensed object from the weak lensing regime continuously to the highly-magnified and distorted strong lensing regime. \href{https://github.com/sibirrer/curved_arcs/blob/v1.0/Notebooks/curved_arc_illustration.ipynb}{\faGithub}}
\label{fig:illustration}
\end{figure*}

Figure~\ref{fig:illustration} illustrates tangential arcs as a function of tangential-to-radial stretch ratio, $\lambda_{\text{tan}}/\lambda_{\text{rad}}$, and curvature. Any other differential or higher order is set to zero in this illustration.
The linear regime is fully described by three parameters in this notation, namely the radial and tangential stretches, and the orientation of one of the eigenvectors, while the curvature is zero ($s_{\rm tan}=0$). The highly non-linear regime requires only the addition of one parameter, namely the curvature along the larger eigenvector (now called tangential stretch). Even a completely round Einstein ring can be fully described in this notation by setting $\lambda_{\text{tan}} = \infty$ with $s_{\rm tan}>0$.
The expression of tangential arcs does allow one to locally describe the lensing phenomenology from the weak lensing to the strong lensing regime.

\subsection{A lens model basis for extended tangential arcs} \label{sec:curved_arc_basis}
To this point in the manuscript, the discussion around tangential arcs have been made at the infinitesimal differential limit. Applications to describe extended arcs require a non-local expression covering the extents of individual arcs or images observed.
We demand the following conditions to be satisfied by the local deflector model around a pre-specified location $\boldsymbol{\theta_0}$, such as the center of an arc:
\begin{enumerate}
	\item The differentials at $\boldsymbol{\theta_0}$ result in $\lambda_{\rm rad}$, $\lambda_{\rm tan}$, $s_{\rm tan}$ and direction $\phi_{\rm tan}$, as specified.
	\item The curvature $s_{\rm tan}$ is constant along the tangential direction, effectively letting the path integral along the tangential eigenvector direction go around in a circle. We denote the radius of this circle as the curvature radius.
	\item Constant tangential stretch $\lambda_{\rm tan}$ on the curvature radius.
	\item Constant radial stretch on the curvature radius.
	\item No curl component on the deflection field.
	\item Zero deflection shift at the location $\boldsymbol{\theta_0}$, $\boldsymbol{\alpha}(\boldsymbol{\theta_0})= 0$.
\end{enumerate}
We identify the following deflector model that uniquely satisfies the criteria mentioned above, of which the deflection angle is given by
\begin{multline} \label{eqn:curved_arc_defl}
 \boldsymbol{\alpha}(\boldsymbol{\theta}) =  s_{\rm tan}^{-1} \left(\frac{\lambda_{\rm tan}- \lambda_{\rm rad}}{\lambda_{\rm rad}\lambda_{\rm tan}} \right) \left( \frac{\boldsymbol{\theta} - \boldsymbol{\theta_c}}{\left| \boldsymbol{\theta} - \boldsymbol{\theta_c}\right|}  -
\frac{\boldsymbol{\theta_0} - \boldsymbol{\theta_c}}{\left| \boldsymbol{\theta_0} - \boldsymbol{\theta_c}\right|} 
  \right) \\
  + \left(1 - \lambda_{\rm rad}^{-1} \right) \left( \boldsymbol{\theta} - \boldsymbol{\theta_0}\right),
\end{multline}
with $\boldsymbol{\theta}_c$ is the centroid position of the curvature radius
\begin{equation}
	\boldsymbol{\theta}_c = \boldsymbol{\theta}_0 - s_{\rm tan}^{-1} \boldsymbol{e}_{\rm rad}.
\end{equation}

Equivalently, the deflector model above can be expressed as a singular isothermal sphere model (SIS) in combination with an MST as
\begin{equation}
	\boldsymbol{\alpha}(\boldsymbol{\theta}) = \lambda_{\rm MST} \left[\boldsymbol{\alpha}_{\rm SIS}(\boldsymbol{\theta}) - \boldsymbol{\alpha}_{\rm SIS}(\boldsymbol{\theta}_0) \right] + \left(1 - \lambda_{\rm MST} \right)  \left( \boldsymbol{\theta} - \boldsymbol{\theta_0} \right),
\end{equation}
with $\lambda_{\rm MST} = \lambda_{\rm rad}^{-1}$ and
\begin{equation}
	\boldsymbol{\alpha}_{\rm SIS}(\boldsymbol{\theta}) = \theta_{\rm E} \frac{\boldsymbol{\theta} - \boldsymbol{\theta_c}}{\left| \boldsymbol{\theta} - \boldsymbol{\theta_c}\right|},
\end{equation}
with Einstein radius 
\begin{equation}
	\theta_{\rm E} = s_{\rm tan}^{-1} \left(1- \frac{\lambda_{\rm rad}}{\lambda_{\rm tan}} \right).
\end{equation}
The centroid matches the curvature radius, the Einstein radius is adjusted such as to matches the ratio of tangential-to-radial stretch ratio, $\lambda_{\rm tan}/\lambda_{\rm rad}$, and the MST term matches the inverse of the radial stretch $\lambda_{\rm rad}$.

We emphasize that this expression is only valid locally, such as around an image of an arc, and is not meant to cover an entire deflection field with multiple images. We refer to Section \ref{sec:application} where we use a local tangential arc parameterization basis separately on multiple images to constrain more complex global deflector models.

\section{Observational invariances} \label{sec:invariances}
Having introduced the formalism of tangentially curved deflectors in describing curved arcs, it is essential to understand and characterize lensing invariances and assumptions for extracting general lensing constraints. We thus dedicate this section to lensing degeneracies and their invariances in the characterization of curved arcs within the locally tangential curved deflector model.
In Section~\ref{sec:general_invariances} we formulate the general class of lensing invariances in an operator notation. We then discuss the specific class of the MST in Section~\ref{sec:mst} and how this degeneracy translates to constraints of curved arcs. In Section~\ref{sec:shape_noise} we discuss shape noise degeneracies in the regime of curved arcs.

\subsection{Operator notation of general lensing invariances} \label{sec:general_invariances}

To characterize general lensing degeneracies inherent in gravitational lensing, we define the following notation:
$\mathbf {L}$ is the lensing operator distorting the source, effectively mapping the lensed coordinates, $\boldsymbol{\theta}$, to the coordinates prior to lensing, $\boldsymbol{\beta}$. In general terms, $\mathbf {L}$ describes a coordinate mapping. The lens equation (Eqn.~\ref{eqn:lens_equation}) can be written in this notation as $\mathbf {L}(\boldsymbol{\theta}) = \boldsymbol{\theta} - \boldsymbol{\alpha}(\boldsymbol{\theta})$.
Given an intrinsic source morphology $\mathbf {S}$, such that $\mathbf {S}(\boldsymbol{\beta})$ describes the intrinsic surface brightness at position $\boldsymbol{\beta}$, the distorted image, $\mathbf {D}(\boldsymbol{\theta})$, can be written as
\begin{equation}
	\mathbf {D}(\boldsymbol{\theta}) = \mathbf {S}(\mathbf{L}(\boldsymbol{\theta})).
\end{equation}
In terms of the operator notation, $\mathbf {L}$ is acting on $\mathbf {S}$ resulting in $\mathbf {D}$, stating as 
\begin{equation} \label{eqn:operator_lens_equation}
	{\mathbf D} = \mathbf {L \circ S}.
\end{equation}

With this notation, we can describe the general invariance between lensing operator $\mathbf {L}$ and source morphology $\mathbf {S}$ resulting in the same image $\mathbf {D}$ with one single additional mapping operator $\mathbf {J}$, by expanding expression (\ref{eqn:operator_lens_equation}) with the unity operator (now written as $\mathbf {J}^{-1}\mathbf {J}$) as
\begin{multline} \label{eqn:degeneracy_general}
	{\mathbf D} = \mathbf {L \circ S} = \mathbf {L \circ} \mathbb{1} \mathbf {\circ S} 
	= \mathbf {L \circ  (J^{-1}  J) \circ S} \\
	= \mathbf {(L \circ  J^{-1})\circ  (J \circ S)} \equiv \mathbf{\tilde{L} \circ \tilde{S}}.
\end{multline}
In the last line, we defined the transformed deflection operator, $\mathbf {\tilde{L}} \equiv \mathbf {L \circ J^{-1}}$, and transformed source, $\mathbf {\tilde{S}} \equiv \mathbf {J \circ S}$, resulting in the same image $\mathbf {D}$.
The only formal requirement on $\mathbf {J}$ in Equation (\ref{eqn:degeneracy_general}) above is that the mapping is bijective and the inverse $\mathbf {J}^{-1}$ is uniquely defined over the extent of image $\mathbf {D}$.

In summary, for any bijective angular mapping operator $\mathbf {J}$, there exists an alternative solution to the lens equation,  ${\mathbf D} = \mathbf{\tilde{L} \circ \tilde{S}}$ with source $\mathbf {\tilde{S}} = \mathbf {J \circ S}$ and lens $\mathbf {\tilde{L}} = \mathbf {L \circ J^{-1}}$. This statement is an operator formulation of the Source Position Transform (SPT) \citep{Schneider:2014, Unruh:2017, Wertz:2018}.

In the presence of two or more images, $\mathbf{D_i}$, $\mathbf{D_j}$, the relative operator translating one image into another can be determined without the knowledge of the intrinsic source $\mathbf{S}$:
\begin{equation}
	\mathbf {D_j} = \mathbf {L_j \circ S} =  \mathbf {L_j \circ L_i^{-1} \circ D_i} = \mathbf {\tilde{L}_j \circ \tilde{L}_i^{-1} \circ D_i}.
\end{equation}
In short, the measurable quantity in lensing under the full consideration of the SPT is the relative distortion operator $\mathbf {L_j \circ L_i^{-1}}$ between two images of the same source\footnote{We refer to \cite{Tessore:2017} for the explicit notation of this invariance in the linear regime of a matix with shear and convergence.}.

In the following, we sequence the general degeneracy operator $\mathbf{J}$ in a scalar component, $\lambda_{\rm MST}$, a linear distortion component, $\mathbf{\Gamma}$, and a third component $\mathbf{O}$ encapsulating any higher order components not captured in the previous two components, as
\begin{equation} \label{eqn:spt_split}
	\mathbf{J} \equiv \lambda_{\rm MST} \mathbf{\Gamma O}.
\end{equation}
The scalar component in this transform is the special case of the MST. The shear component $\mathbf{\Gamma}$ characterizes the shape noise, while the non-linear component $\mathbf{O}$ characterizes any higher order distortion of the SPT. 

In Section~\ref{sec:mst} we further discuss the MST component within the framework of curved arcs, and in Section~\ref{sec:shape_noise} we discuss the shape noise aspects of the SPT, with a brief discussion on higher order terms.

\subsection{Mass-sheet transform (MST)} \label{sec:mst}
The mass-sheet transform (MST) is the scalar component of the general SPT (Eqn.~\ref{eqn:spt_split}). This scalar component is a multiplicative transform of the lens equation (Eqn.~\ref{eqn:lens_equation}) preserving image positions (and thus any higher order differentials too) under a linear source displacement $\boldsymbol{\beta} \rightarrow \lambda\boldsymbol{\beta}$ and was introduced by \cite{Falco:1985no, Gorenstein:1988} as such

\begin{equation} \label{eqn:lens_equation_MST}
  \lambda_{\rm MST} \boldsymbol{\beta} = \boldsymbol{\theta} - \lambda_{\rm MST} \boldsymbol{\alpha}(\boldsymbol{\theta}) - (1 - \lambda_{\rm MST}) \boldsymbol{\theta}.
\end{equation}
The term $(1 - \lambda_{\rm MST}) \boldsymbol{\theta}$ in the equation above describes an infinite sheet of convergence (or mass) and hence the name mass-sheet transform.
The corresponding transform of the convergence profile is given by
\begin{equation} \label{eqn:mst}
  \kappa_{\rm MST}(\boldsymbol{\theta}) =  \lambda_{\rm MST} \kappa(\boldsymbol{\theta}) - (1 - \lambda_{\rm MST}).
\end{equation}

The MST can be described as a global transform of the convergence and hence it can lead to physical solutions for a wide range of values of $\lambda_{\rm MST}$. A fact that makes the MST a prominent and relevant degeneracy for many applications, in particular the measurement of the Hubble constant with time-delay cosmography \citep[e.g.,][]{Schneider:2013pm, Birrer:2016zy, Sonnenfeld:2018, Kochanek:2020a, Blum:2020, Birrer:2020}.
Only observables related to the absolute source size, intrinsic magnification of the lensed source, the absolute lensing potential, or the relative time delay when imposing a known cosmology with absolute distances, are able to break this degeneracies.

The differentials of the lens equation (e.g. Jacobian \ref{eqn:jacobian} and flexion \ref{eqn:flexion_differentials}) transform under an MST as
\begin{equation} \label{eqn:differentials_MST}
  \mathbf{A'} = \lambda_{\rm MST} \mathbf{A} \quad\text{,}\quad \mathbf{D'} = \lambda_{\rm MST} \mathbf{D}.
\end{equation}
The coefficients in the Jacobian and higher order differentials are not constrained by imaging observables unless other constraints or assumptions on the lensing profile are inserted and thus do not serve themselves to be observables. 

Equivalently, the MST scales the radial and tangential eigenvectors as
\begin{equation}
	\lambda'_{\rm tan} = \lambda^{-1}_{\rm MST} \lambda_{\rm tan} \quad\text{,}\quad \lambda'_{\rm rad} = \lambda^{-1}_{\rm MST} \lambda_{\rm rad}.
\end{equation}
The quantities that remain locally invariant under the MST is the ratio of tangential to radial eigenvalue $\lambda_{\text{tan}}/ \lambda_{\text{rad}}$ that describes the relative distortions and any directional quantities (eigenvector direction)\footnote{This is equivalent to the reduced shear expression.}. Considering the third order derivatives, the curvature $s_{\rm tan}$ and $s_{\rm rad}$ remain invariant under the MST. The derivatives of the eigenvalues follow the same scaling with the MST as the eigenvalues themselves.

\subsection{Shape degeneracies in curved arcs} \label{sec:shape_noise}
Beyond the MST, the remaining aspects of a linear distortion are the reduced shear components ($\mathbf{\Gamma}$ in Eqn.~\ref{eqn:spt_split}). These components are changing the ellipticity of the intrinsic source.
In the regime where the lensing operator $\mathbf{L}$ is linear, any linear SPT, $\mathbf{\Gamma}$, leads to a linear transform of $\mathbf{\tilde{L}} = \mathbf{L \circ \Gamma^{-1}}$ and is thus indistinguishable from the reduced shear. This degeneracy is generally known as shape noise \citep[see e.g.][]{Bernstein:2002}. We refer to Appendix \ref{app:ellipticity_shear} for a shear and intrinsic shape notation convenient in transforming according to a linear SPT.
However, if the lensing operator $\mathbf{L}$ is non-linear, such as in the regime of curved arcs, the shape noise transformed lensing operator couples the differentials non-linearly and can give rise to a curl component in the deflection operator $\mathbf{\tilde{L}}$.

\begin{figure*}
  \centering
  \includegraphics[angle=0, width=130mm]{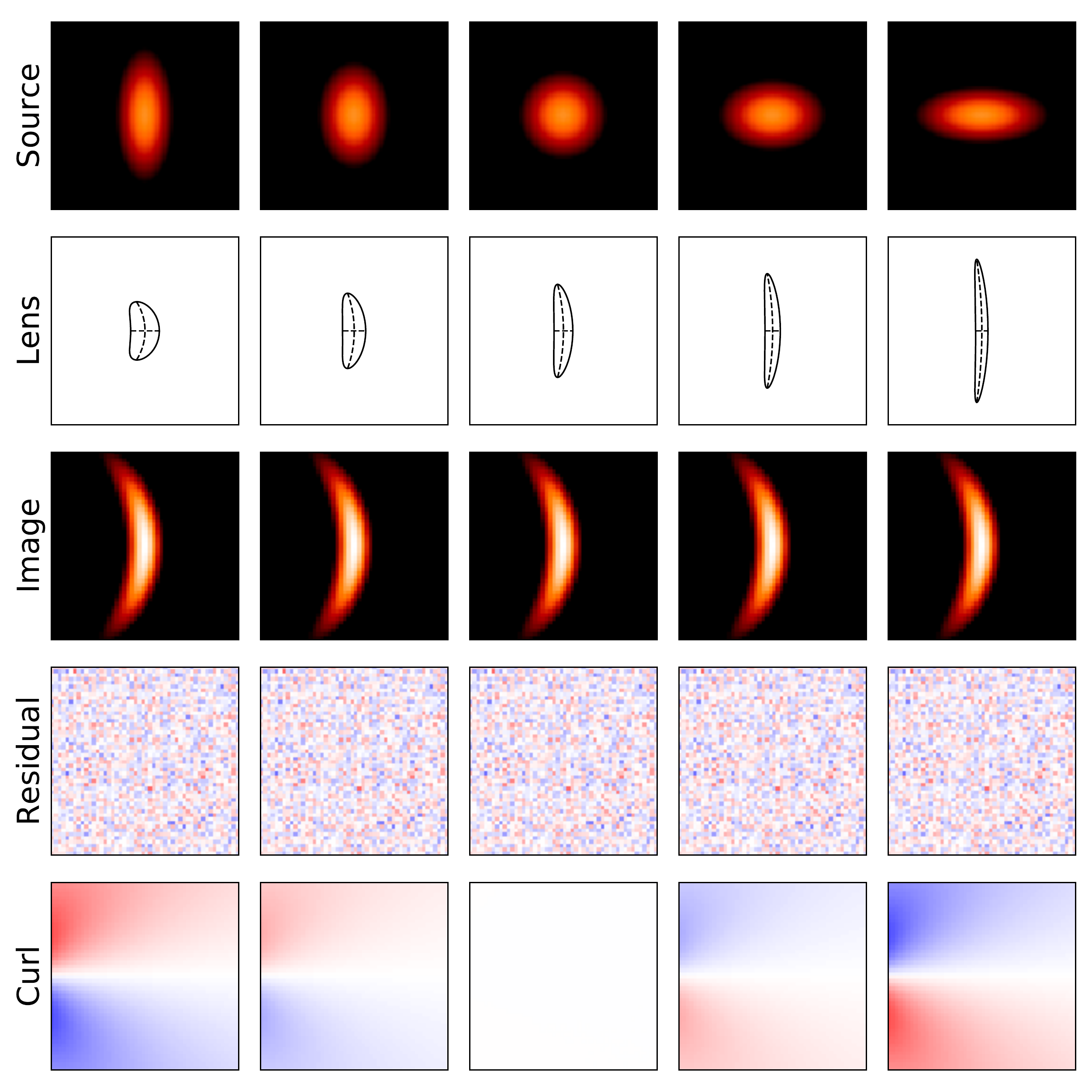}
  \caption{Demonstration of the shape noise component of the SPT applied on a curved arc on-axis relative to tangential eigenvector. The middle column corresponds to a reference example of a round intrinsic source (top) being distorted by a curl-free curved arc (Eqn.~\ref{eqn:curved_arc_defl}, second row) resulting in the lensed curved arc (third row). The re-fitting with an SPT mapping leads, by construction, to a perfect fit (fourth row indicates reduced residuals of the fit) without curl (fifth row). The other columns correspond to an enforced different elliptical shape of the intrinsic source ($\mathbf{\tilde{S}}$) with a lensing operator ($\mathbf{\tilde{L}}$) to perfectly describe the SPT. The resulting fit to the data is perfect but the required curl is non-zero. The off-axis distortions with the SPT is presented in Figure~\ref{fig:degeneracy_spt_off_axis}.
  \href{https://github.com/sibirrer/curved_arcs/blob/v1.0/Notebooks/spt_curved_arc.ipynb}{\faGithub}}
\label{fig:degeneracy_spt_on_axis}
\end{figure*}

\begin{figure*}
  \centering
  \includegraphics[angle=0, width=130mm]{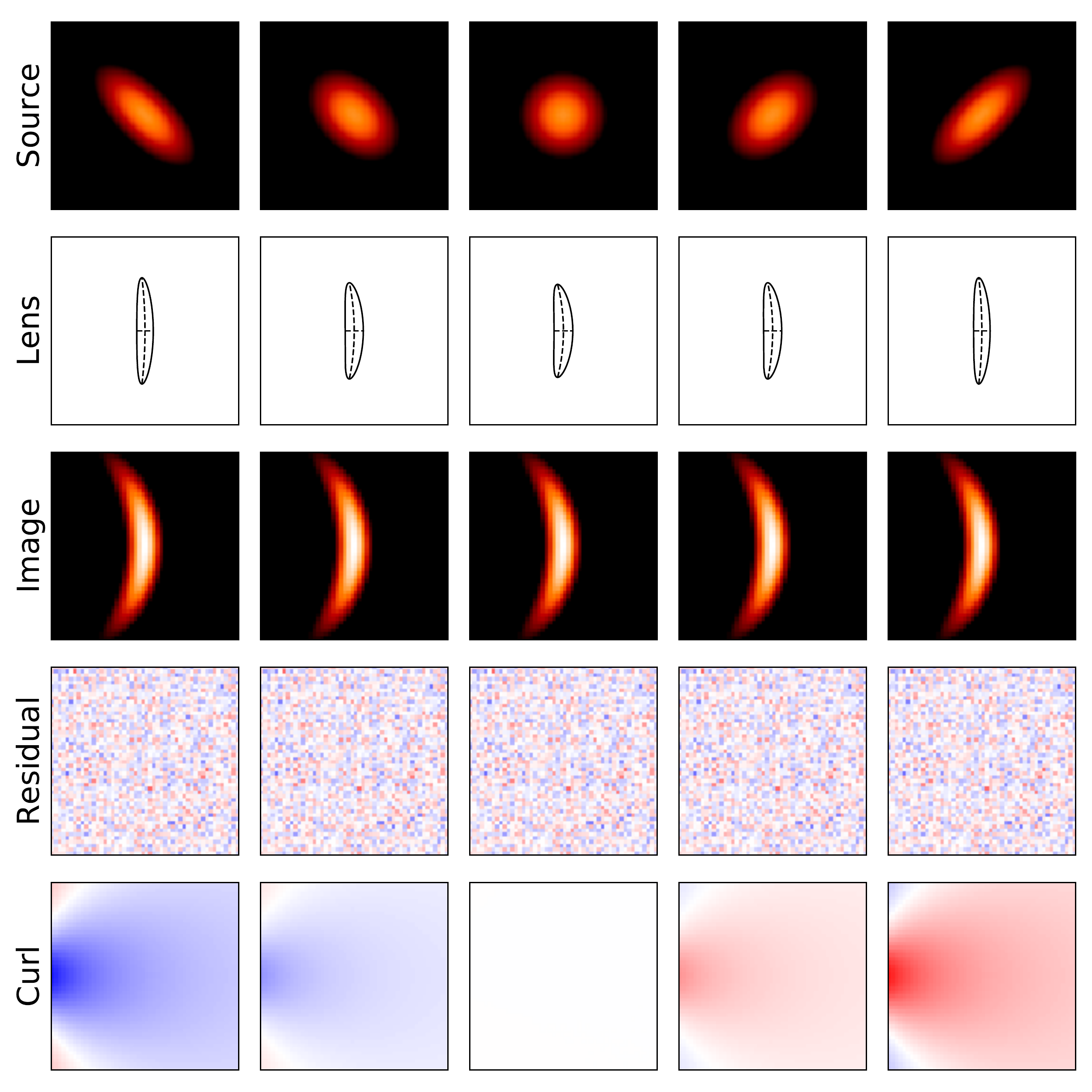}
  \caption{Demonstration of the shape noise component of the SPT applied on a curved arc off-axis relative to tangential eigenvector. The middle column corresponds to a reference example of a round intrinsic source (top) being distorted by a curl-free curved arc (Eqn.~\ref{eqn:curved_arc_defl}, second row) resulting in the lensed curved arc (third row). The re-fitting with an SPT mapping leads, by construction, to a perfect fit (fourth row indicates reduced residuals of the fit) without curl (fifth row). The other columns correspond to an enforced different elliptical shape of the intrinsic source ($\mathbf{\tilde{S}}$) with a lensing operator ($\mathbf{\tilde{L}}$) to perfectly describe the SPT. The resulting fit to the data is perfect but the required curl is non-zero. The on-axis distortions with the SPT is presented in Figure~\ref{fig:degeneracy_spt_on_axis}.
  Off-axis shape distortions are better constraint by curved arcs than on-axis distortions as illustrated in the difference in the residuals between this figure and Figure~\ref{fig:degeneracy_spt_on_axis}.
  \href{https://github.com/sibirrer/curved_arcs/blob/v1.0//Notebooks/spt_curved_arc.ipynb}{\faGithub}}
\label{fig:degeneracy_spt_off_axis}
\end{figure*}

We illustrate the non-linear coupling by performing an SPT on a curved arc with a round source and an extended curved tangential deflector model given by expression \ref{eqn:curved_arc_defl}. In Figure~\ref{fig:degeneracy_spt_on_axis} the shear transform is performed along the tangential axis and in Figure~\ref{fig:degeneracy_spt_off_axis} the transform is performed along the orthogonal shear modulus. By construction, the SPT results in a perfect match of the original arc for all cases. For the on-axis SPT (Figure~\ref{fig:degeneracy_spt_on_axis}), the local eigenvectors and tangential curvature are transformed by the expected relative tangential and radial size of the source. The extended deflection field, however, contains a significant curl contribution. For the off-axis distortions (Figure~\ref{fig:degeneracy_spt_off_axis}), even at the center of the arc, significant curl contributions arise from the SPT.

\begin{figure*}
  \centering
  \includegraphics[angle=0, width=130mm]{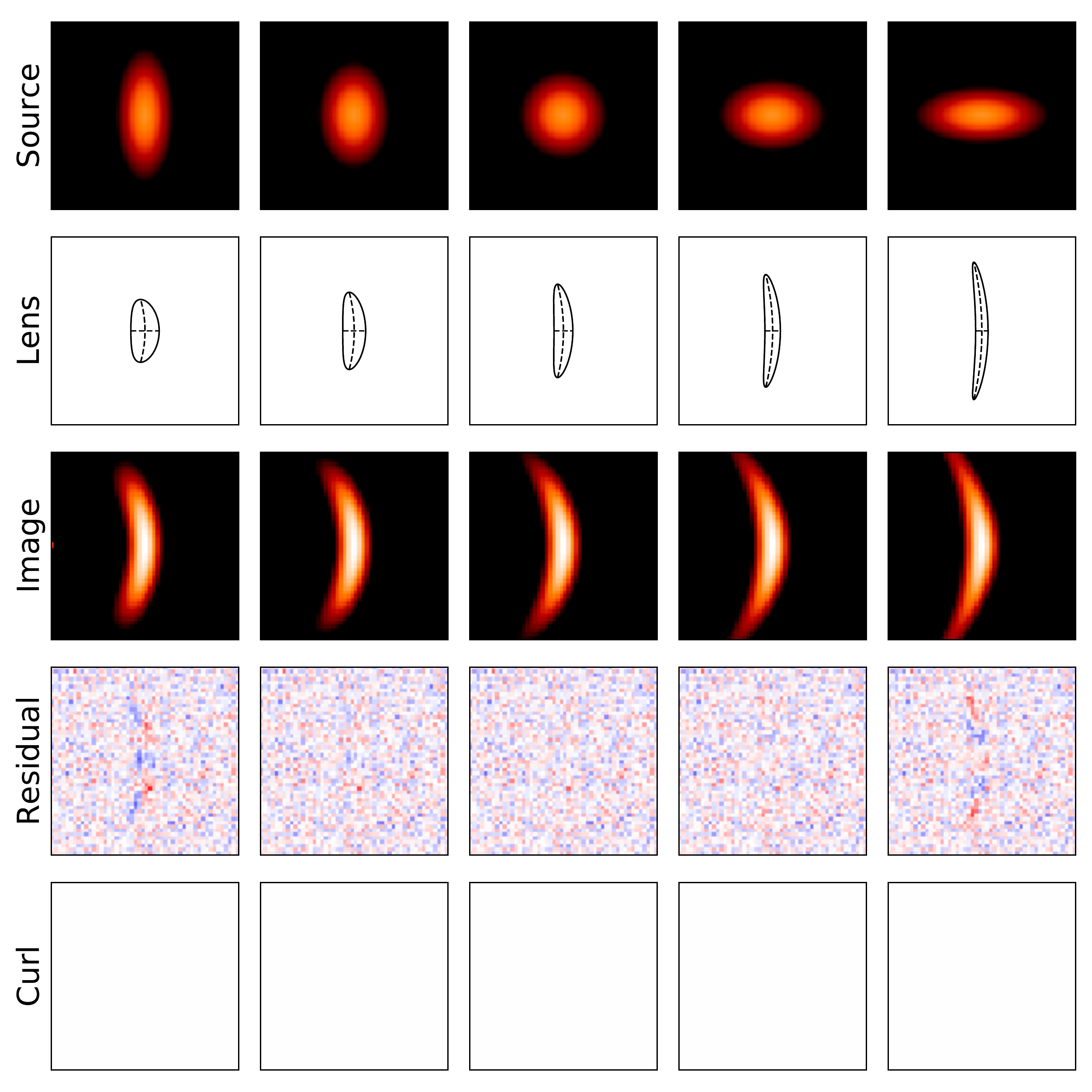}
  \caption{Demonstration of the shape noise component of the curl-free curved arc approximation of the SPT applied on a curved arc on-axis relative to tangential eigenvector. The middle column corresponds to a reference example of a round intrinsic source (top) being distorted by a curl-free curved arc (Eqn.~\ref{eqn:curved_arc_defl}, second row) resulting in the lensed curved arc (third row). The re-fitting with a tangentially curved deflector model best approximating the SPT leads, by construction, to a perfect fit (fourth row indicates reduced residuals of the fit). The tangentially curved deflector models have, by design, no curl components (fifth row). The other columns correspond to an enforced different elliptical shape of the intrinsic source ($\mathbf{\tilde{S}}$) with a lensing operator ($\mathbf{\tilde{L}}$) of a curved arc (Eqn.~\ref{eqn:curved_arc_defl}) approximating the SPT. The resulting fit to the data is not perfect and the enforced curl-free nature of the model leads to distinguishable intrinsic source shape features. The on-axis distortions with the curved arc approximated SPT is presented in Figure~\ref{fig:degeneracy_off_axis}.
  \href{https://github.com/sibirrer/curved_arcs/blob/v1.0//Notebooks/spt_curved_arc.ipynb}{\faGithub}}
\label{fig:degeneracy_on_axis}
\end{figure*}

\begin{figure*}
  \centering
  \includegraphics[angle=0, width=130mm]{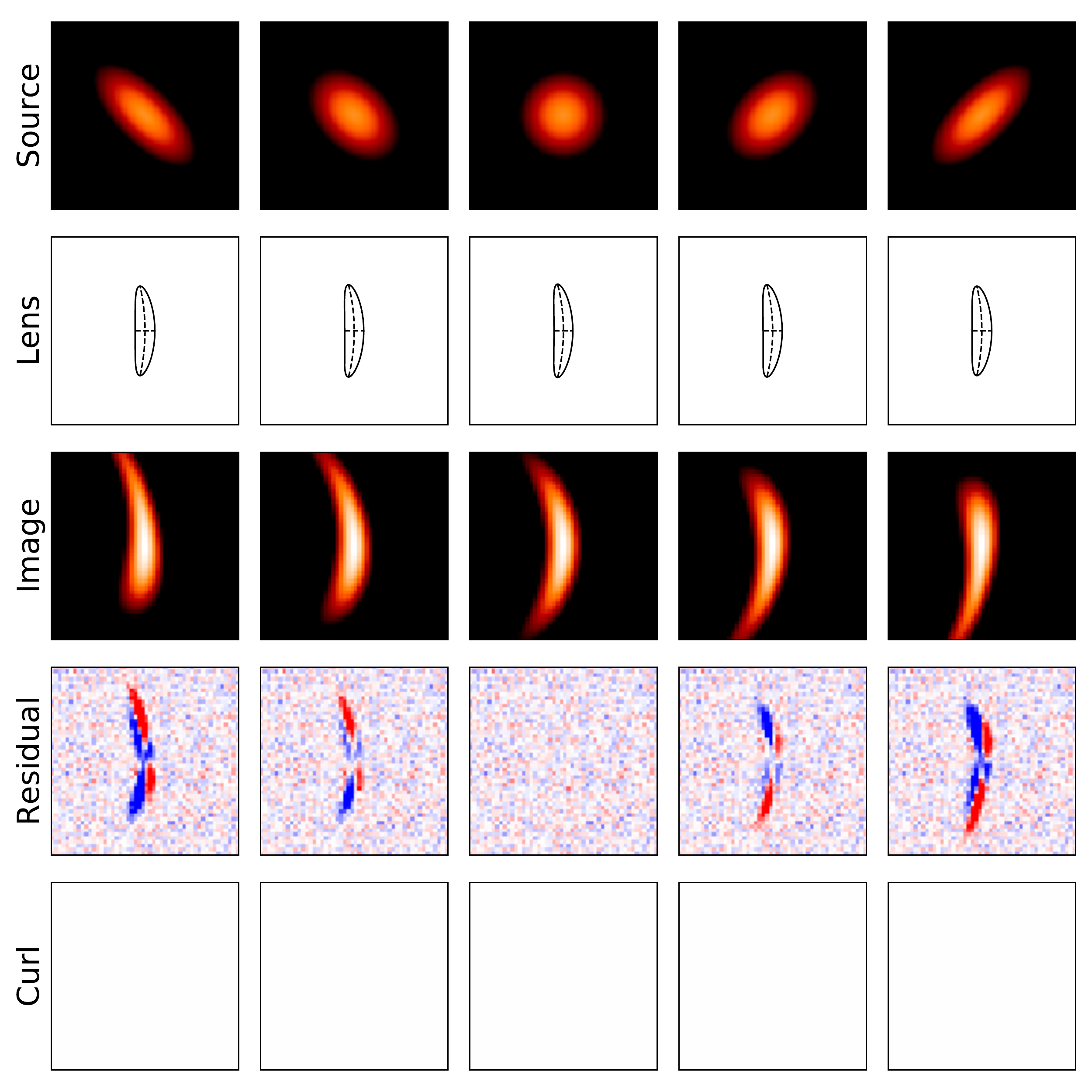}
  \caption{Demonstration of the shape noise component of the curl-free curved arc approximation of the SPT applied on a curved arc off-axis relative to tangential eigenvector. The middle column corresponds to a reference example of a round intrinsic source (top) being distorted by a curl-free curved arc (Eqn.~\ref{eqn:curved_arc_defl}, second row) resulting in the lensed curved arc (third row). The re-fitting with a tangentially curved deflector model best approximating the SPT leads, by construction, to a perfect fit (fourth row indicates reduced residuals of the fit) without curl (fifth row). The curve arc models have, by design, no curl components (fifth row). The other columns correspond to an enforced different elliptical shape of the intrinsic source ($\mathbf{\tilde{S}}$) with a lensing operator ($\mathbf{\tilde{L}}$) of a curved arc (Eqn.~\ref{eqn:curved_arc_defl}) approximating the SPT. The resulting fit to the data is not perfect and the enforced curl-free nature of the model leads to distinguishable intrinsic source shape features. The off-axis distortions with the curved arc approximated SPT is presented in Figure~\ref{fig:degeneracy_on_axis}.
  \href{https://github.com/sibirrer/curved_arcs/blob/v1.0//Notebooks/spt_curved_arc.ipynb}{\faGithub}}
\label{fig:degeneracy_off_axis}
\end{figure*}

In the next approach, we restrict the lensing transform $\mathbf{\tilde{L}}$ to a curl-free curved arc (Eqn.~\ref{eqn:curved_arc_defl}), while demanding the source morphology to be sheared.
Figure~\ref{fig:degeneracy_on_axis} shows the approximate SPT with a curl-free curved arc for on-axis distortions, re-fit to give the best possible fit to the original arc generated with a round source. While the curved arc parameter fit follows the same infinitesimal properties as for the SPT at the center of the arc, residuals in the extent of the arcs remain. Thus, within the assumption of a curl-free tangentially curved deflector model, the shape noise can be constrained. Off-axis distortions, as illustrated in Figure~\ref{fig:degeneracy_off_axis}, are  more constrained, as the remaining residual patterns indicate. This feature can also be linked to the missing curl component in the center of the arc, as expected by the SPT.
The closest approximations to the exact SPT within the curved arc lens model family does not allow us to adequately describe the observed arcs to the signal-to-noise level of the simulation for substantial distortions of the source, thus restricting shape noise.

Higher order SPT components ($\mathbf{O}$ in Eqn.~\ref{eqn:spt_split}) in general lead to source transforms that deviate from elliptical shapes. A subset of these transforms can lead to curl-free mappings $\mathbf{\tilde{L}}$. One mathematically possible case is when there is no lensing ($\mathbf{\tilde{L}} = \mathbb{1}$), then the shape of the source, $\mathbf{\tilde{S}}$ is a curved arc itself. However, the physical plausibility of galaxies resembling in an intrinsic arc-like shape needs to be considered and the likelihood of higher-order morphological shapes can be estimated empirically from the shapes of the entire galaxy population in low lensing environments.
We further refer to \cite{Schneider:2014} for a discussion on higher order SPTs in the axi-symmetric case for global mass distributions, and to \cite{Unruh:2017} for non-axi-symmetric cases.

In this section, we did not discuss the impact of a point spread function (PSF), In the example in Section (\ref{sec:example_mock}) we incorporate a PSF corresponding to a HST observation. We point out that uncertainties in the ellipticity of the PSF can also lead to degeneracies related to shape noise and thus accurate and precise PSF estimates are essential for studies of gravitational lensing, in particular when extracting significant information from individual objects.

We also note that when multiple arcs of the same source are present and the local lensing distortions are simultaneously reconstructed, this will add further constraining power on the SPT components depending on the relative alignment of the different curved arcs. So even if there is a curl-free SPT allowed to reproduce one arc by transforming the source morphology in a particular way, the lensing operator of an additional image may require significant curl components to match the observations.
Given that off-axis shape distortions are better constrained than on-axis components, multiple images of arcs that are asymmetrically aligned, meaning on- and off-axis directions in the individual arcs correspond to different axes in the intrinsic source plane, do suppress the shear-ellipticity degeneracy more efficiently. Fully connected Einstein rings further enhance the suppression of the shear-ellipticity degeneracy.

\section{Constraining global mass distributions} \label{sec:profile_constraints}
Theoretical discussions in the literature in regards to mass profile constraints are primarily using positional constraints and magnification ratios and are often tied and applicable to a specific mass profile family.
In this section, we discuss and illustrate which observational features extractable by curved arcs allow us to constrain what specific aspects of global mass distributions in the non-linear regime of gravitational lensing.
We first discuss the tangential constraints related to ellipticity and external shear of a mass distribution (Section \ref{sec:tang_constraints}) and then in a second step we separately discuss the radial constraints provided by observed curved arcs (Section \ref{sec:rad_constraints}).
This section is accompanied by Appendix \ref{app:pemd} where we state the specific functional form of the global lens models we use in this work as an example.

\subsection{Azimuthal constraints}\label{sec:tang_constraints}

Tangential distortions in strong gravitational lensing imprint signal about the asymmetric mass distribution in the main deflector and along the line of sight. To first order, the asymmetry can be described by an elliptical mass distribution and external shear.
Positional constraints of quadruply imaged sources can only partially break the shear-ellipticity degeneracy \citep[see e.g.,][]{Schechter:2019, Luhtaru:2021} under fixed radial profile constraints. \cite{ORiordan:2020} studies positional and magnification constraints on the joint ellipticity-power law radial slope, not considering degeneracies with external shear.

In the formalism of extended curved arcs, the following shape quantities provide information about the azimuthal structure of the lens: (i) the change in the tangential stretch along the azimuth of the deflector, $\partial_t \lambda_{\rm tan}$, (ii) change in the curvature direction along the azimuthal direction, (iii) change in the curvature radius along the azimuthal direction.

\begin{figure*}
  \centering
  \includegraphics[angle=0, width=150mm]{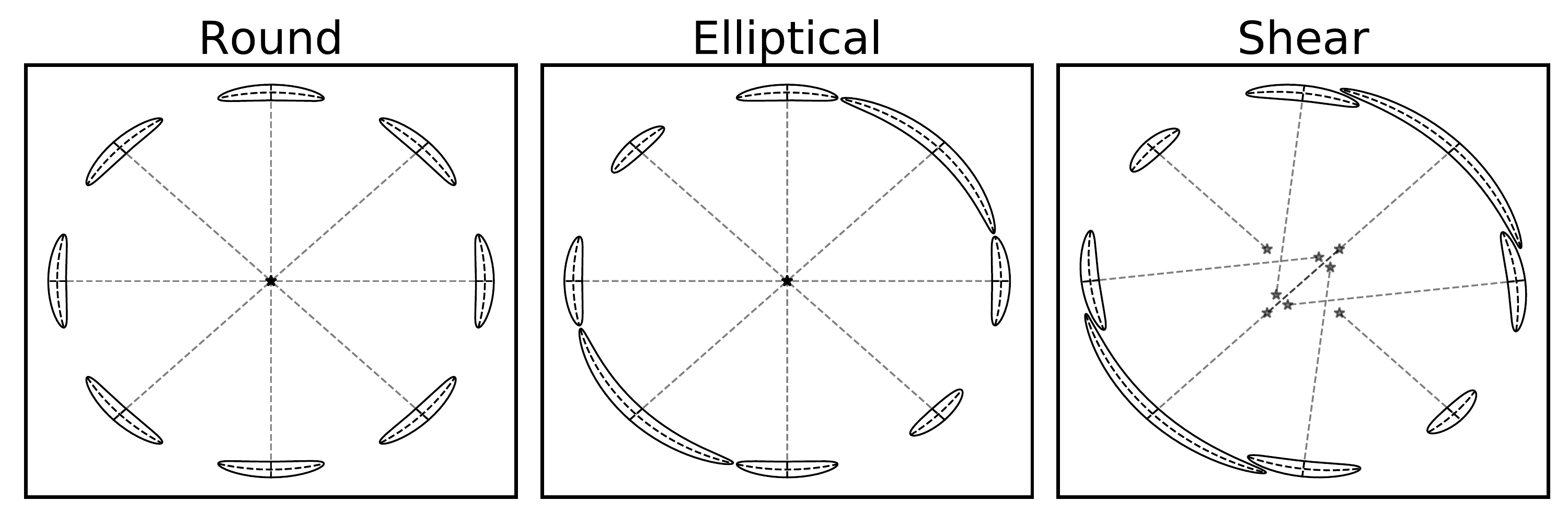}
  \caption{Illustration of curved arc properties at a fixed radial distance along the azimuthal axis for three different lens models. Left: Round lens model, resulting in a fully symmetric appearance of arcs. Middle: elliptical mass distribution, causing a change in the tangential stretch $\partial_t\lambda_{\rm tan}$ along the azimuth with a $180^{\circ}$ symmetry imposed by the lens model symmetry. The curvature radius and direction, however, remain centered as it is the case for a round mass distribution. Right: Round mass density with an addition of an external shear component. While the change in the tangential stretch varies almost identically as for the case of an elliptical mass distribution, the additional unambiguous feature of the shear component is the fact that the direction  of the curvature in the arc is offset from the mass distribution center with an altered curvature radius.
  \href{https://github.com/sibirrer/curved_arcs/blob/v1.0/Notebooks/curved_arc_illustration.ipynb}{\faGithub}}
\label{fig:azimuthal_constraints}
\end{figure*}

Figure~\ref{fig:azimuthal_constraints} illustrates the curved arc properties at a fixed radial distance along the azimuthal axis for three different lens models. The round model exhibits, imposed by its symmetry, identical curved arc structure along the azimuth with the curvature radius and direction pointing towards the center of the deflector profile. The elliptical mass model, here described as a power-law elliptical mass distribution (PEMD, see Appendix \ref{app:pemd} for details), causes a change in the tangential stretch $\lambda_{\rm tan}$ along the azimuth with a $180^{\circ}$ symmetry imposed by the lens model symmetry. The curvature radius and direction, however, remain centered on the deflector mass. In the third case, we illustrate the azimuthal behavior of a round mass density with an addition of an external shear component. While the change in the tangential stretch varies almost identically as for the case of an elliptical mass distribution, the additional unambiguous feature of the shear component is the fact that the direction of the curvature in the arc is offset from the deflector center with an altered curvature radius.

The example illustrated in Figure~\ref{fig:azimuthal_constraints} demonstrate how extended resolved arcs are able to break the ellipticity-shear degeneracy. The formalism of curved arcs is able to capture these constraints. We do not discuss azimuthal structure beyond a dipole and external shear but expect that the curved arc formalism and approach is also able to effectively describe and present observational signatures in more complex regimes of azimuthal structure\footnote{see e.g. a study with multi-pole moments and their impact on Hubble constant measurements by Van de Vyvere et al. in prep}.

\subsection{Radial constraints} \label{sec:rad_constraints}

The primary radial constraint from gravitational lensing of a mass profile is the Einstein radius $\theta_{\rm E}$. In the round case, the Einstein radius marks the radius where the tangential stretch $\lambda_{\rm tan}$ diverges and changes its sign, known as the critical curve. The next-order leading term characterizing the radial profile is the radial stretch eigenvalue $\lambda_{\rm rad}$. This value, however, is not an observable due to the MST and only ratios of eigenvalues are observable. The leading order measurable quantity by gravitational lensing observables is the normalized differential radial stretch of $\partial_r\lambda_{\rm rad}/\lambda_{\rm rad}$ measured as the average finite differential between two arcs at different radial distance from the critical curve.
The quantity $\partial_r\lambda_{\rm rad}/\lambda_{\rm rad}$ can be equivalently expressed as radial derivatives of the deflection angle $\boldsymbol{\alpha}$ or the lensing potential $\psi$ (Eqn.~\ref{eqn:potential})
\begin{equation}\label{eqn:radial_constraints_analogue}
	\frac{\partial_r\lambda_{\rm rad}}{\lambda_{\rm rad}} = \frac{\alpha^{''}}{1 - \alpha'}
	= \frac{\psi^{'''}}{1 - \psi^{''}},
\end{equation}
where $^{'}$ denotes the radial derivative\footnote{We also refer to \cite{Sonnenfeld:2018} to the use and derivation of the right hand side of Equation \ref{eqn:radial_constraints_analogue}.}.

The invariant quantity at the Einstein radius when the radial differential is scaled relative to the Einstein radius, is given by 
\begin{equation}\label{eqn:xi_inv}
	\xi_{\rm rad} \equiv \theta_{\rm E} \frac{ \partial_r\lambda_{\rm rad}(\theta_{\rm E})}{\lambda_{\rm rad}(\theta_{\rm E})}.
\end{equation}
We note that the quantity $\xi_{\rm rad}$ is effectively equivalent in the constraining power to the expression introduced by \cite{Kochanek:2020a} 
\begin{equation}
 \theta_{\rm E} \frac{ \partial_r\lambda_{\rm rad}(\theta_{\rm E})}{\lambda_{\rm rad}(\theta_{\rm E})} \propto \theta_{\rm E} \frac{\alpha''(\theta_{\rm E})}{(1-\kappa_{\rm E})},
\end{equation}
where $\kappa_{\rm E}$ is the convergence at the Einstein radius.
The only difference between the expression in this work and by \cite{Kochanek:2020a} is the representation of the MST, either by the absolute radial stretch eigenvector or the convergence at the Einstein radius, respectively.
Both expressions allow for a model-independent interpretation and translation of lensing constraints from one mass-profile family to another. This relation has been used, for example, by \cite{Shajib:2021} to derive constraints on a family of more flexible mass density profiles based on original constraints derived with power-law density profiles.

In the following, we discuss what aspects of curved arcs allow us to constrain $\partial_r\lambda_{\rm rad}/\lambda_{\rm rad}$. We identified three distinct aspects; (i) relative arc thickness measurements, (ii) relation of $\partial_r\lambda_{\rm rad}/\lambda_{\rm rad}$ to tangential stretch due to underlying symmetries, and (iii) positional constraints of arcs. 

\subsubsection{Differential radial thickness of arcs}
The most direct constraints on the radial differentials can be made by measuring the relative thickness of multiply imaged arcs appearing at different radial distances from the critical curve. This measurement is demanding, as arcs are usually not stretched along the radial direction ($\lambda_{\rm rad} \approx 1$) and thus thin. Relative thickness differences of a few percent are often below the resolution of the instrument. 
We emphasize that radial differential thickness, though the most intuitive constraining aspect, is often not the dominating constraining factor in the inference of radial differentials but instead subdominant to the aspects mentioned in the following paragraphs.

\subsubsection{Differential tangential extent of arcs}
Differentials in the tangential extent of arcs do also allow us to constrain the radial differentials when imposing symmetries between the differential quantities.
Specifically, an azimuthally symmetric deflection field obeys the following relation between tangential stretch and relative source and image position in radial direction:
\begin{equation} \label{eqn:sherical_tan_constraints}
\lambda_{\rm tan} = \frac{\theta_{r}}{\beta_{r}}.
\end{equation}
This relation is simply reflecting the fact that when rotating the source position around the center of the deflector, the image positions are demanded to rotate with the same angle.
This symmetry argument leads to an imposed relation between the differential of the tangential stretch in radial direction, $d\lambda_{\rm tan}/dr$, and the radial eigenvector $\lambda_{\rm rad}$.
In particular, differentiating relation (\ref{eqn:sherical_tan_constraints}) along the radial direction results in
\begin{equation} \label{eqn:rad_tan_sym}
	\partial_r\lambda_{\rm tan} = \frac{1}{\beta_r} - \frac{\theta_r}{\beta_r^2}\frac{d\beta_r}{d\theta_r}
	= \frac{\lambda_{\rm tan}}{\theta_r} \left(1 - \frac{\lambda_{\rm tan}}{\lambda_{\rm rad}} \right),
\end{equation}
where in the last equality above we substituted $\beta_r = \theta_r/\lambda_{\rm tan}$ (Eqn.~\ref{eqn:sherical_tan_constraints}) and $d\theta_r / d\beta_r = \lambda_{\rm rad}$.
A version of the MST invariant relation of expression (\ref{eqn:rad_tan_sym}) reads
\begin{equation} \label{eqn:rad_tang_symmetry}
	\frac{\partial_r\lambda_{\rm tan}}{\lambda_{\rm tan}} 
	= \frac{1}{\theta_r} \left(1 - \frac{\lambda_{\rm tan}}{\lambda_{\rm rad}} \right).
\end{equation}
Imposing this relation allows one to derive constraints on the radial density profile while utilizing measurements of tangential stretch differences. Relative tangential stretch differences are often easier to measure as the extent of the arc is larger in the tangential direction, well beyond the seeing limit.

\begin{figure*}
  \centering
  \includegraphics[angle=0, width=170mm]{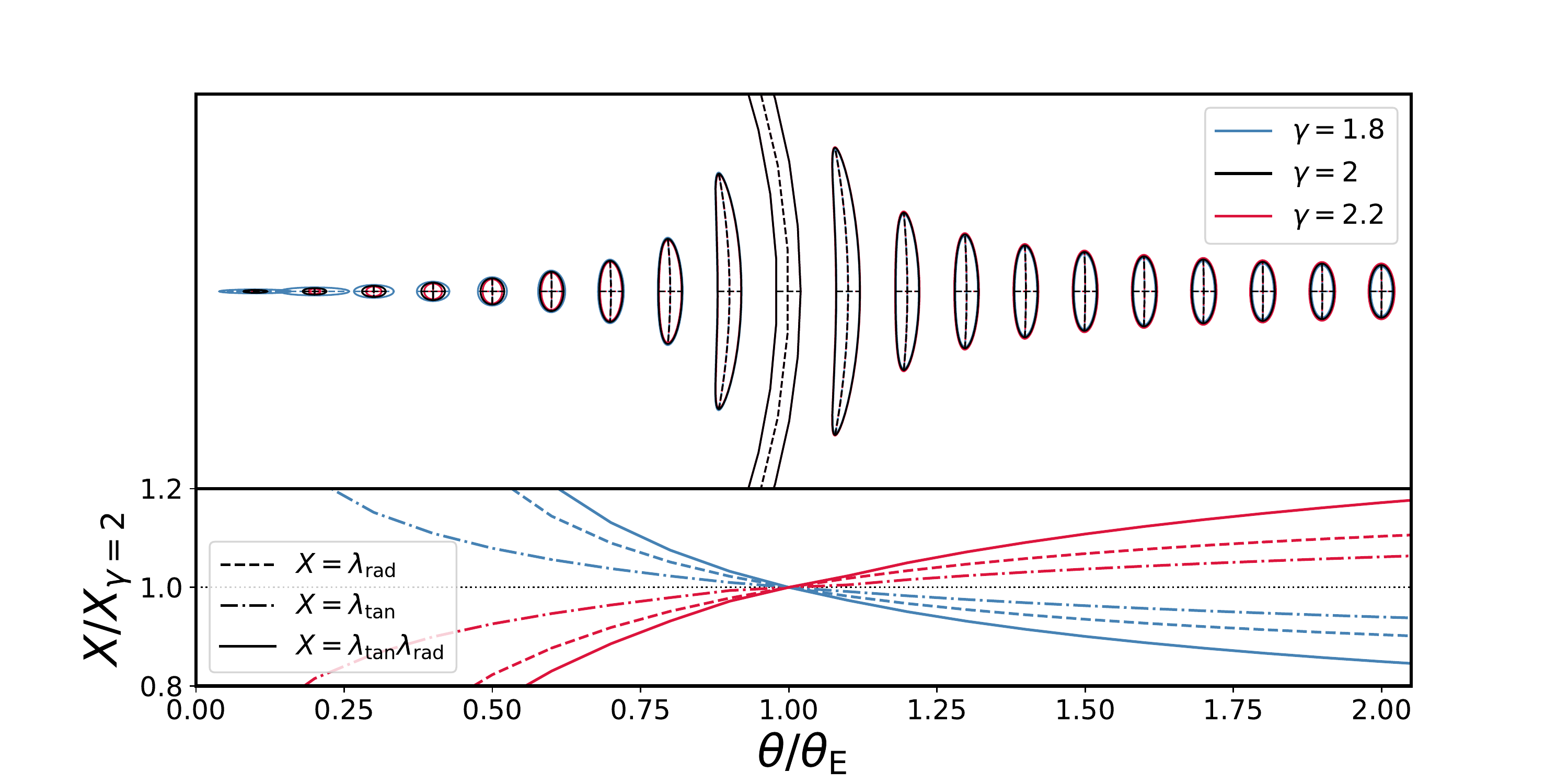}
  \caption{Illustration of the differences of tangential arcs relative to the scale at the Einstein radius for three different values of the power-law slope of a power-law mass profile, as specified by the colors in the legend. Top: Curved arcs at different radii for a fixed intrinsic source size normalized to match the width at the Einstein radius. Bottom: Difference in the tangential (dotted-dashed), radial (dashed) and magnification (solid) of the arcs relative to the isothermal density profile (black). The differentiability between different constant power-law slopes is provided in both, relative radial stretch, and relative tangential stretch. Positional constraints on the appearance of multiple images are not part of this figure and are covered in Figure~\ref{fig:radial_point_source}.
  \href{https://github.com/sibirrer/curved_arcs/blob/v1.0/Notebooks/curved_arc_illustration.ipynb}{\faGithub}}
\label{fig:radial_constraints}
\end{figure*}

In Figure~\ref{fig:radial_constraints} we illustrate the differences of tangential arcs relative to the scale at the Einstein radius for three different values of the power-law slope of a constant power-law mass profile. The differentiability between different power-law slopes is provided in relative radial stretch and relative tangential stretch.

We emphasize that the relative tangential stretch relation along specific directions can also be caused by azimuthal structure. Specific assumptions, such as the absence or presence of an azimuthal twist as a function of radius may impact radial constraints on the profile, if they are primarily derived from the tangential scale ratio, a statement also made by \cite{Kochanek:2021}.

\subsubsection{Positional constraints}

Positional constraints of image pairs of the same source also contain information about the radial differential stretch $\partial_r\lambda_{\rm rad}/ \lambda_{\rm rad}$.
In this section we discuss the round deflector case where two magnified images appear, one inside the Einstein radius, $\theta_{\rm in}$, and one outside the Einstein radius, $\theta_{\rm out}$. A third de-magnified solution of the lens equation is at or very close to the center of the deflector density and we ignore this image in this discussion, as it is often unobserved.

Image pairs satisfy the lens equation (Eqn.~\ref{eqn:lens_equation}). The lens equation (Eqn.~\ref{eqn:lens_equation}) for the two solutions $\theta_{\rm in}$ and $\theta_{\rm out}$ arising from the same source position $\beta$ demands that
\begin{equation}
	\theta_{\rm in} - \alpha(\theta_{\rm in}) = \beta = \theta_{\rm out} - \alpha(\theta_{\rm out}).
\end{equation}

To investigate radial dependences on the relative solution of the lens equation, we expand the solution relative to the Einstein radius, where the solution is given by $\theta_{\rm E} - \alpha(\theta_{\rm E}) = 0$. We can write the solution of the lens equation in an integral form of the source displacement from the origin as
\begin{equation}
	\int_{\theta_{\rm E}}^{\theta_{\rm in}} \frac{d\beta(\theta')}{d\theta'}d\theta' = \beta = \int_{\theta_{\rm E}}^{\theta_{\rm out}} \frac{d\beta(\theta')}{d\theta'}d\theta'.
\end{equation}

Defining the relative radial distance from the Einstein radius for the two images as $\Delta \theta_{\rm in} \equiv \theta_{\rm E} - \theta_{\rm in}$ and $\Delta \theta_{\rm out} \equiv \theta_{\rm out} - \theta_{\rm E}$ and noting that $d\beta_r(\theta)/d\theta_r = \lambda^{-1}_{\rm rad}(\theta)$, we can write the radial solution of the lens equation as 
\begin{equation} \label{eqn:lens_eqn_integral}
	\int_{0}^{\Delta \theta_{\rm in}} \frac{1}{\lambda_{\rm rad}(\theta_{\rm E} - \theta')} d\theta' =  \int_{0}^{\Delta \theta_{\rm out}} \frac{1}{\lambda_{\rm rad}(\theta_{\rm E} + \theta')} d\theta'.
\end{equation}
Writing $\lambda_{\rm rad}(\theta)^{-1}$ as a Taylor expansion around $\theta_{\rm E}$ and only considering first and second order terms in $\Delta \theta$, Equation~\ref{eqn:lens_eqn_integral} can be approximated by
\begin{equation} \label{eqn:position_asymetry_quadratic}
	\frac{1}{\lambda_{\rm rad}}\Delta \theta_{\rm in}  + \frac{1}{2}\frac{\partial_r\lambda_{\rm rad} }{\lambda_{\rm rad}^2} \Delta \theta_{\rm in}^2 \approx 
	\frac{1}{\lambda_{\rm rad}}\Delta \theta_{\rm out}  - \frac{1}{2}\frac{\partial_r\lambda_{\rm rad} }{\lambda_{\rm rad}^2} \Delta \theta_{\rm out}^2.
\end{equation}
We further simplify the expression above explicitly stating the asymmetry in the image appearance $\Delta \theta_{\rm out}/\Delta \theta_{\rm in}$ as a function of the mean displacement of the image pair relative to the critical curve, $(\Delta\theta_{\rm out} + \Delta\theta_{\rm in})/2$. Approximating $\Delta \theta_{\rm out}^2 /\Delta \theta_{\rm in} \approx \Delta \theta_{\rm out}$, Expression \ref{eqn:position_asymetry_quadratic} can be expressed by
\begin{equation} \label{eqn:position_asymetry_approx}
	\frac{\Delta \theta_{\rm out}}{\Delta\theta_{\rm in}} \approx 1 + \frac{\partial_r\lambda_{\rm rad} }{\lambda_{\rm rad}} \frac{\Delta \theta_{\rm in} + \Delta \theta_{\rm out}}{2}.
\end{equation}

This relation shows that the radial asymmetry in the appearance of images relative to the Einstein radius (or in more general terms the critical curve) is directly linked to the reduced derivative of the radial stretch, $\partial_r\lambda_{\rm rad}/ \lambda_{\rm rad}$ at the Einstein radius, and is linear as a function of mean radial separation. Relation \ref{eqn:position_asymetry_approx} is effectively equivalent to the relation presented by \cite{Sonnenfeld:2018} expressed in terms of differentials of the lensing potential.

\begin{figure}
  \centering
  \includegraphics[angle=0, width=80mm]{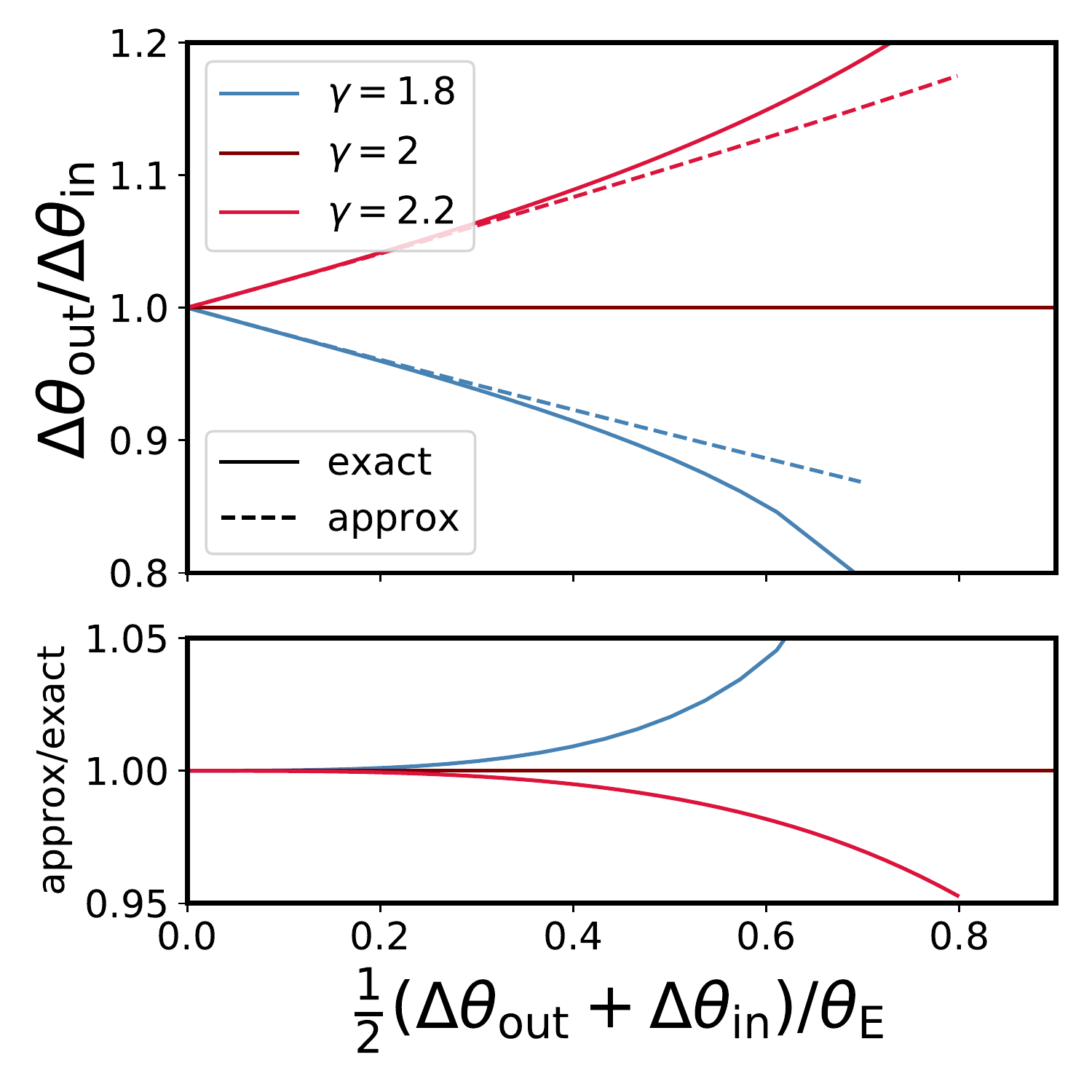}
  \caption{Ratio of relative radial distances of an image pair relative to the Einstein radius for different radial power-law density slopes. $\Delta \theta_{\rm out}$ is the distance from the outer image to the Einstein radius and $\Delta \theta_{\rm in}$ is the distance from the inner image to the Einstein radius. Top panel: Distance ratio of $\Delta \theta_{\rm out} / \Delta \theta_{\rm in}$ as a function of average distance normalized to the Einstein radius, $1/2 (\Delta \theta_{\rm out} + \Delta \theta_{\rm in}) / \theta_{\rm E}$. Solid lines indicate the exact solution of the lens equation, while dashed lines show the approximated linear solution ignoring terms beyond $\partial_r\lambda_{\rm rad}/ \lambda_{\rm rad}$ given by Equation~\ref{eqn:position_asymetry_approx}. Lower panel: Ratio of exact to approximate pair asymmetry. Imprints of the distortions of extended sources are illustrated in Figure~\ref{fig:radial_constraints}.
  \href{https://github.com/sibirrer/curved_arcs/blob/v1.0/Notebooks/curved_arc_illustration.ipynb}{\faGithub}}
\label{fig:radial_point_source}
\end{figure}

To investigate the validity of the approximation in expression (\ref{eqn:position_asymetry_approx}), we compare in Figure~\ref{fig:radial_point_source} the relative radial image position for different slopes of a power-law radial density profile. For the isothermal density profile ($\gamma = 2$), $\lambda_{\rm rad}$ is constant and the exact solution as well as the approximation predicts an exact symmetry in the image pair appearance. For shallower and steeper slopes $\partial_r\lambda_{\rm rad}$ as well as higher order terms are non-zero and an asymmetry in the appearance is observed. The approximate solution proves to be accurate to one percent in the inferred power-law slope out to about $0.4 \times \theta_{\rm E}$ in mean separation of the images.

\section{Example and discussions} \label{sec:application}

In the previous sections, we have introduced the formalism to describe local curved arcs and have elaborated lensing degeneracies and constraints from a theoretical point of view.
The goal of this section is to outline potential practical applications and outline extensions. We provide an example of deriving macro-model independent lensing constraints from a multiply imaged extended source in Section~\ref{sec:example_mock}.
In Section~\ref{sec:example_science} we provide suggestions in the usage of the presented formalism for different science cases, and in Section~\ref{sec:limitations} we discuss limitations and possible extensions of the current formalism.

\subsection{Example: Model-independent extraction of lensing information of a quadruply imaged extended source} \label{sec:example_mock}

Here we provide an example of utilizing the curved arc formalism to derive macro-model independent constraints on the deflector model for a quadruply imaged extended source. We assess this alternative to fitting a global deflector model and discuss what constraints are data-driven and what constraints are model-driven.

\subsubsection{Model set up and fitting procedure}
Our input deflector model is a PEMD profile with a circular Gaussian source. We are using a Hubble Space Telescope typical point spread function (PSF) width (as a Gaussian kernel), pixel scale and noise level (Figure~\ref{fig:example_reconstruction} top left).
We explicitly chose an example of a macro-model that can not be represented globally by the degrees of freedom we allow for with individual tangentially curved deflector models.

For the model fitting, we define four regions of the image that capture the individual distorted images and chose four independent extended tangentially curved deflector models in the reconstruction process. The local deflector models have the parameterization of the tangential and radial stretch eigenvalues, $\lambda_{\rm tan}$ and $\lambda_{\rm rad}$, the direction $\phi_{\rm rad}$, tangential curvature $s_{\rm tan}$, and a tangential eigenvalue differential $\partial_t \lambda_{\rm tan}$. The underlying deflector model is stated in Appendix~\ref{app:sie_arc}. In addition to the distortions, each deflector model patch has two additional uniform deflection displacement parameters that effectively map the center of the curved arc to the center of the intrinsic source and contain the positional information. Per curved arc, there are seven free parameters.
For the source morphology, we allow for a free ellipticity, as parameterized with the eccentricity moduli (see Appendix~\ref{app:ellipticity}). This description ensures a full exploration of the shape-noise degeneracy discussed in Section \ref{sec:shape_noise}. We fix the intrinsic source size for the purpose of an efficient sampling and the fact that the MST adds an additional full degeneracy in the overall scales of the inferred eigenvalues (see Section~\ref{sec:mst}).
We use \textsc{lenstronomy} in the \texttt{joint-linear} mode, meaning that the likelihood of the different patches and different deflector models are evaluating given the same source morphology surface brightness amplitude. This mode has been used by \cite{Yang:2020, Yang:2021} to reconstruct the intrinsic sources of multiply lensed galaxies in the cluster lensing environment. PSF and noise properties are matched to the input simulation during the inference.

In total, the sampling contains 30 non-linear parameters. For the parameter posterior sampling we follow \cite{Birrer2015_basis_set}. We first find a maximum likelihood position using a Particle Swarm Optimizer \citep[PSO;][]{pso} exploring a large volume of parameter space (200 particles for up to 500 iterations). We then use the obtained best-fit value as a starting point with significantly narrower proposal distribution to perform a Monte Carlo Markov Chain (MCMC) using \textsc{emcee} \citep{emcee} (with 300 particles for 2000 burn-in and 2000 sampling iterations to ensure convergence of the chain).

\subsubsection{Model-independent curved arc constraints}

\begin{figure*}
  \centering
  \includegraphics[angle=0, width=170mm]{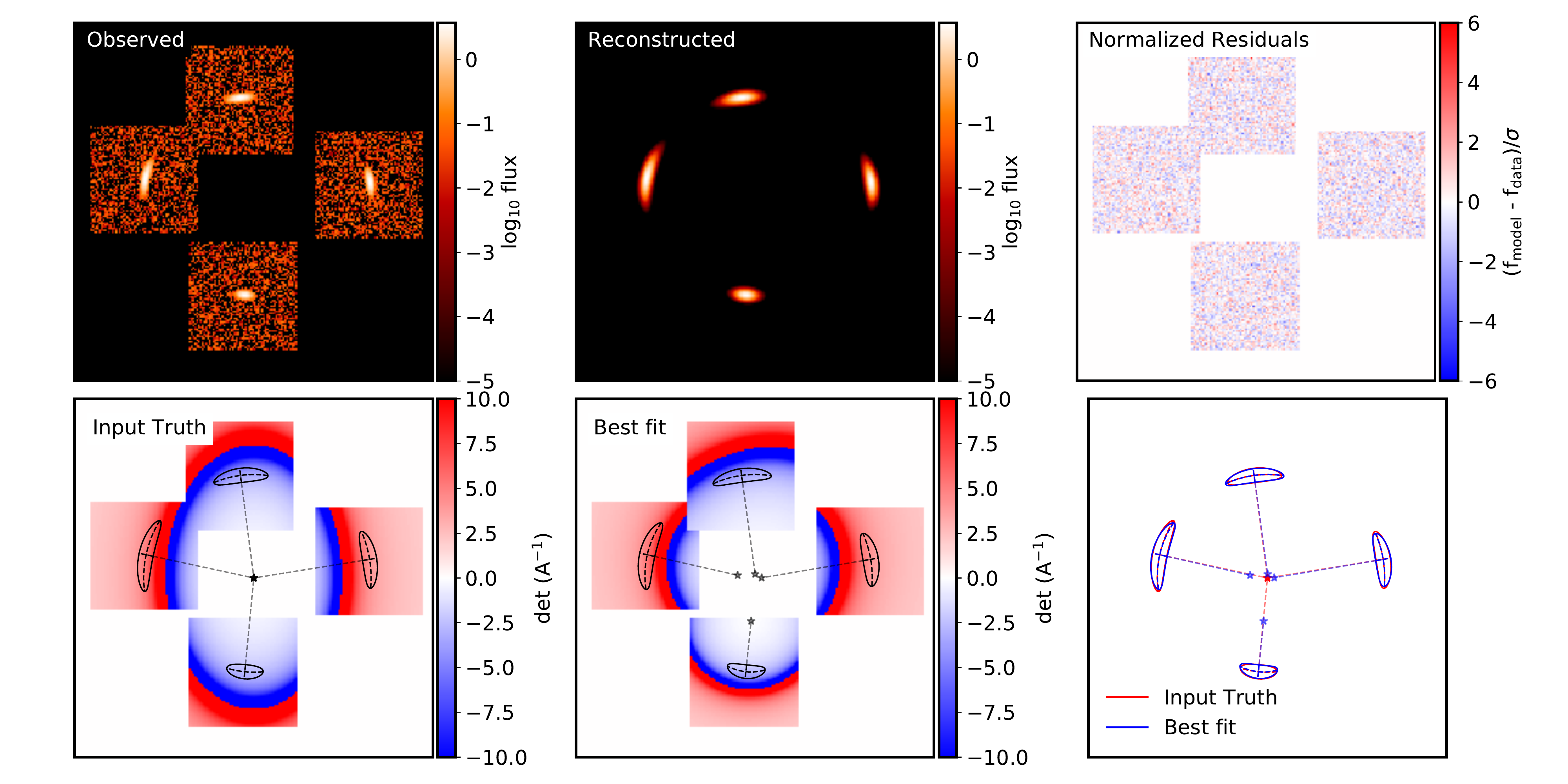}
  \caption{Example of applying curved arc description in deriving macro-model independent constraints on the lens model. The input mock data is generated with a PEMD mass profile (top left: input mock data, lower left: local curved arc differentials from the input truth at the arc positions and magnification map). The individual arcs are fit within separate cut-out regions with independent tangentially curved deflector models. Only the source is demanded to share the same morphology among the different curved arcs. The local curved deflector formalism allows us to describe the input data to the noise level (middle top: best fit reconstruction of the arcs, top right: reduced residuals of the model - data). The lensing constraints derived from the local tangentially curved deflector models at the position of the arcs are accurate (bottom right). The set of local tangentially curved deflector models do not require to fully describe the global macro model in regions absent of data constraints (bottom middle: local magnification predictions).
  \href{https://github.com/sibirrer/curved_arcs/blob/v1.0/Notebooks/local_vs_global_fit.ipynb}{\faGithub}}
  \label{fig:example_reconstruction}
\end{figure*}

Figure \ref{fig:example_reconstruction} presents the best fit reconstruction using the curved arc formalism. The local curved arc deflectors centered at the appearances of the arcs reproduce the observables to the noise level of the input data, without relying on specific assumptions on the functional form of the global macro lens model. Thus, we expect from this modeling procedure an accurate extraction of the lensing information independent of the underlying global deflector properties.

Beyond the best fit, the posteriors on the curved arc parameters capture effectively the lensing information in the extended data that go beyond the positional information. The blue contours in Figure~\ref{fig:example_posteriors} correspond to the model-independent inference of the tangential and radial stretch eigenvectors ($\lambda_{\rm tan}$, $\lambda_{\rm rad}$) at the positions of the images (subscript 0-3) of the example displayed in Figure \ref{fig:example_reconstruction}. In addition to the eigenvalues, we also show the eccentricity moduli of the source shape ($e_1$ and $e_2$,  see Appendix \ref{app:ellipticity}). Not displayed are the direction $\phi_{\rm rad}$, curvature $s_{\rm tan}$ and tangential differential $\partial_t \lambda_{\rm tan}$ parameters for the individual local tangentially curved deflector models. The sampling is performed under flat priors in the parameters stated. The posteriors are consistent with the input truth (black line, evaluated from the input macro-model at the positions of the curved arcs).

We notice a significant degeneracy between the intrinsic shape parameter $e_1$ and the eigenvalues of all the local arcs. The direction of $e_1$ corresponds to horizontal and vertical distortions and are almost on-axis with the tangential direction of all the four images. We showed in Section~\ref{sec:shape_noise} that the shape noise is less well constrained on-axis to the tangential arc than off-axis (comparison of e.g. the residuals of Figure~\ref{fig:degeneracy_on_axis} for on-axis and Figure~\ref{fig:degeneracy_off_axis} for off-axis shape noise). Thus, we expect a stronger breaking in the off-axis direction of the shape noise ($e_2$ in this example), than in on-axis direction ($e_1$ in this example). On-axis shape noise is also degenerate with the tangential-to-radial stretch ratio (e.g. Figure~\ref{fig:degeneracy_on_axis}). The degeneracies and relative uncertainties in this example are a reflection and confirmation of the discussion presented in Section~\ref{sec:shape_noise}.

\subsubsection{Global model-dependent constraints}
We can compare the constraints on the same quantities as measured by the curved arc inference when performing an inference on a global deflector model and then evaluating the local quantities from the global posterior model. 

\begin{figure*}
  \centering
  \includegraphics[angle=0, width=170mm]{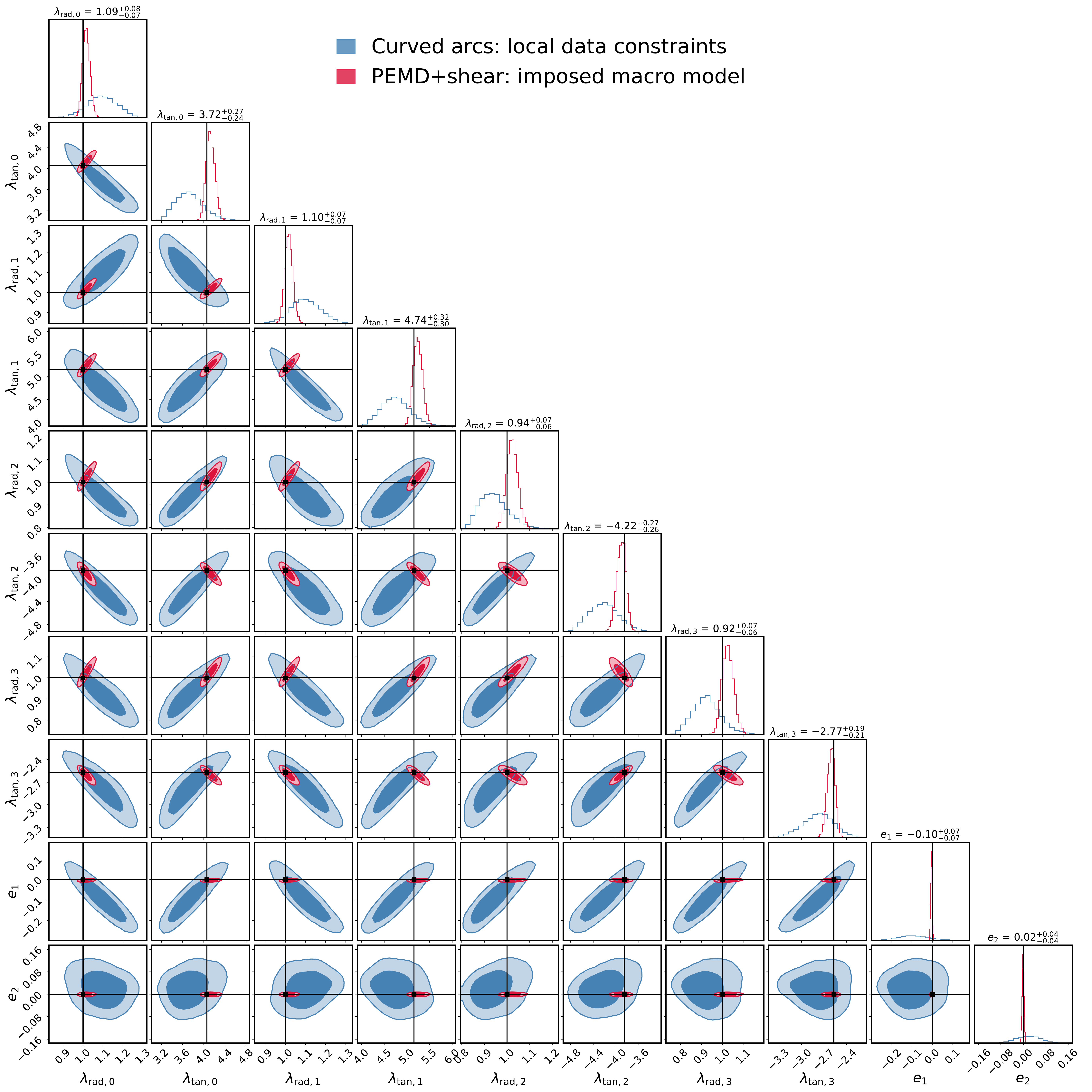}
  \caption{Comparison of a model-independent and model-dependent inference of the tangential and radial stretch eigenvectors ($\lambda_{\rm tan}$, $\lambda_{\rm rad}$) at the positions of the images (subscript 0-3). The example input and fit with curved arcs is shown in Figure~\ref{fig:example_reconstruction}. $e_1$ and $e_2$ correspond to the intrinsic source eccentricity moduli (see Appendix~\ref{app:ellipticity}).
  \textbf{Blue contours:} Posteriors of the extended local arc inference. The different images are constrained by independent curved arc parameters. Only the intrinsic source is demanded to be identical when predicting the individual arc surface brightness. The source size is held fixed, thus the posteriors reflect a slice within the MST (i.e. re-scaling all eigenvalues results in an equally valid model with re-scaled source). Not shown are the direction $\phi_{\rm rad}$, curvature $s_{\rm tan}$ and tangential differential $\partial_t \lambda_{\rm tan}$ parameters for the individual local tangentially curved deflector models. Uncertainties quoted in the figures correspond to these blue contours.
  \textbf{Red contours:} Post-processed predictions of the same quantities derived from a global PEMD+shear model inference of the same data. The intrinsic source size was a free parameter. The assumptions on the chosen global mass profile breaks the MST.
 \textbf{Black lines:} Truth input values computed from the input lens model (PEMD+shear).
  The eigenvalues (modulo an overall scaling) of the blue contours can be considered as a measurement provided by the data. No continuity in the deflection field between the curved arc locations is required. The additional constraints between blue and red contours do solely come from the specific imposed global model assumptions, in addition to the MST breaking in the PEMD+shear scenario.
  Accuracy in the red contours is only guaranteed if the chosen lens model assumptions are valid.
  \href{https://github.com/sibirrer/curved_arcs/blob/v1.0/Notebooks/local_vs_global_fit.ipynb}{\faGithub}
  }
  \label{fig:example_posteriors}
\end{figure*}

In our example, we chose as a global macro-model as an elliptical power-law mass density (PEMD, Appendix~\ref{app:pemd}) model with external shear (Appendix~\ref{app:shear}) with flat priors on all the parameters. The PEMD+shear model is a popular model of choice in many applications on galaxy-scale strong gravitational lensing modeling applications. In addition to the source shape parameters, we also allow the source size parameter to vary in this scenario to be agnostic to MST breaking effects.
The red contours in Figure~\ref{fig:example_posteriors} correspond to the post-processed posterior predictions from the global PEMD+shear model inference of the same data for the same quantities as derived for the curved arc inference.
The differences in the posterior widths between the curved arc measurements and a global lens model inference is attributed to the specific assumptions imposed by the choice of the macro-model parameterization and the translation of the prior space.
Global mass profile assumptions can be discussed in terms of required radial and tangential symmetries demanded by a certain model (see Section~\ref{sec:profile_constraints}). In tangential direction, the PEMD+shear model allows only specific configurations of the curvature direction and strength, and the tangential differentials along the azimuthal direction (Section~\ref{sec:tang_constraints}). In addition, the symmetry requires a specific relation of tangential and radial stretch (Eqn.~\ref{eqn:rad_tang_symmetry}).
The asymmetry in the appearance of the images further allows to add constraints on the relative radial stretch differential $\partial_r \lambda_{\rm rad} /\lambda_{\rm rad}$ beyond the explicitly measured differential width in radial direction reflected in the curved arc posteriors.
These assumptions and symmetry considerations allows the imposed model to break the shape noise and the related degeneracies present in the curved arc inference.

Furthermore, the PEMD+shear model imposes a one-to-one relation between the measurable quantity $\partial_r \lambda_{\rm rad} /\lambda_{\rm rad}$ and the power-law slope (Eqn.~\ref{eqn:pl_constraints}). This assumption imposed by the model effectively breaks the MST and simultaneously allows the model to constrain the source size.

The parameterization we chose inherently contains the input truth and, thus, allows for an accurate recovery of the input quantities. Had we chosen a different parameterization of the macro-model, the general expectation is that the posteriors are within the margins of the curved arc measurement, modulo an overall MST re-scaling not represented in the displayed curved arc posteriors, to be consistent with the data. However, any narrowing of the posterior due to further implied constraints on the macro model might lead to biases within the boundaries of the curved arc posterior.

We will discuss certain aspects of this example in Section~(\ref{sec:limitations}) in more broader terms in light of possible applications and limitations.

\subsection{Science cases} \label{sec:example_science}
In this section we highlight several science cases where our formalism may find beneficiary applications. A more uniform approach to quantifying lensing constraints across different scientific studies and analyses may also result in an overall better ability to utilizing constraints obtained for originally addressing a specific science questions and then translating the constraints to other investigations.

\subsubsection{Dark Matter: Locally resolved vs unresolved small-scale distortions}
Unresolved flux ratio statistics of multiply imaged quasars is a powerful probe of small scale dark matter clustering and constraining the nature of dark matter \citep{Dalal:2002, Hsueh:2020, Gilman:2020wdm, Gilman:2020mc}. Interpretations of the flux ratios require reference flux ratios predicted by a smooth macro model.
Current flux ratio statistics constraints are derived from quadruply lensed quasars only \citep{Hsueh:2020, Gilman:2020wdm} and the positional constraints of the images and the deflector light are the primary sources of information to establish a macro-model reference prediction.
Assessments of potential systematics in regard to an assumed macro-model parameterization have been studied by \cite{Hsueh:2016, Gilman:2017, Hsueh:2017, Hsueh:2018} and is a potential source of noise.

The curved arc formalism, applied in similar way as for the example in Section~\ref{sec:example_mock}, allows one to establish a reference local flux-ratio prediction based on the extended host galaxy, without relying on assumptions on the macro model. The example in Section~\ref{sec:example_mock} translates in a  1\% flux ratio prediction, below the current measurement errors of the fluxes \citep{Nierenberg:2020}, thus making the statistical not limited by macro-model uncertainties. On one hand, such an approach requires sufficient host galaxy light components around the quasars, potentially restricting such an analysis to a subset of the quadruply lensed quasar systems. On the other hand, this approach can be also employed in using doubly lensed quasar, a much larger population of lenses, for systems with extended host information equivalently.

In the fully resolved regime of extended arcs in the absence of quasars, a perturbative description of extended arc might be able to replace global model fitting in characterizing the abundances of small scale structure in the lens and along the line of sight, as done in the literature \citep[e.g.,][]{Vegetti:2012, Hezaveh:2016, Birrer:2017}. Substructure signal is generally an anomaly of required local lensing perturbations to match the appearances of multiply imaged sources with a single, yet inherently unknown, morphological structure of the source.

\subsubsection{Time-delay cosmography and Hubble constant measurement}
The relevant radial quantity to derive from the mass density profile to achieve an accurate time-delay prediction is the local convergence at the Einstein radius, which is not a direct observable from lensing data \citep{Kochanek:2002}.
We recommend to derive solely invariant quantities from modeling imaging data, i.e. as quantified in expression \ref{eqn:xi_inv} on the radial profile. In a second step, one can translate these constraints with additional data, such as kinematics. Due to the tight coupling between radial and tangential constraints, careful assessment of the tangential structure assumptions need to be taken as well \citep{Kochanek:2021}.
The Fermat potential prediction can then be re-scaled by a factor of the relative local convergence at the Einstein radius between the initial model used in extracting the lensing information, and the one constrained by external data. A special case of such an analysis is presented by \cite{Birrer:2020} in using the most direct parameterization relevant for the time-delay prediction, the MST itself, in translating constraints from the PEMD models to a more general form of mass density profiles constrained by stellar kinematics observations.

Physically interpretable mass models can be well approximated by a pure MST within a range exceeding 10\% in the MST \citep{Kochanek:2020a, Blum:2020, Birrer:2020}. Higher-order radial differentials can potentially distinguish variations among the families of models but are hard to measure in practice.
We refer to Section 2 of \cite{Birrer:2020} for a detailed discussion of data constraints and physical descriptions of density profiles following approximately an MST relative to a baseline model.

\subsubsection{Large scale structure and the statistics of gravitational lenses}

Searches for strong gravitational lenses in current and ongoing large area imaging surveys, such as the Dark Energy Survey (DES) and the Hyper-Supreme-Cam survey (HSC) have resulted in hundreds of promising galaxy-galaxy scale candidate lenses \citep[see e.g.,][]{Jacobs:2019, Sonnenfeld:2018hsc}. With the next generation large area ground and space based surveys (Vera Rubin Observatory LSST, Euclid, Nancy Grace Roman Space Telescope), of order $10^5$ galaxy-galaxy lenses will be discovered \citep[e.g.,][]{Collett:2015a}. The number of curved arcs, where non-linear curvature can be detected, even in the absence of a detectable counter image, may well be up to an order of magnitude larger, simply by the argument of lensing cross-section.

%The advantage of reduced shape noise in the strong lensing regime relative to the linear lensing regime and image multiplicity, combined with the expected number of curved arcs, contain significant information about the galaxy-halo connection from cluster down to galaxy scales.
The advantage of reduced shape noise in the strong lensing regime relative to the linear lensing regime and image multiplicity, combined with the expected number of curved arcs, is that we gain significant information about the galaxy-halo connection from cluster down to galaxy scales.
Proposed statistical studies on the radial density profiles of galaxies using positional and magnification information \cite[see e.g.,][]{Blandford:1987, Kochanek:1987, Sonnenfeld:2021} can be enhanced with the full information encompassed in curved arcs. Strong lenses may also be able to provide significant cosmic shear information \citep{Birrer:2017los, Birrer:2018_cosmic_shear, Kuhn:2020, Fleury:2021}, potentially even in tomographic mode. These are only two specific examples utilizing partial information contained in the non-linear lensing observables.

The description introduced in this work may also help with simulations and calibrations of large scale weak lensing surveys. In particular, in investigations into next-to-leading order lensing effects and potential systematics in the shape measurements as a cause \citep[see][for such a discussion in regard to flexion]{Schneider:2008} may be needed for the next generation weak lensing surveys.

All in all, a continuous formalism to describe observables from the weak to the strong lensing regime allows one to self-consistently combine the currently distinct cosmological probes gaining synergies and complementarity in systematics and constraints.

\subsection{Discussion of limitations and extensions} \label{sec:limitations}
The focus of this work is primarily to present a framework and methodology to allow the science investigator to assess impacts of certain assumptions on specific science cases and to translate constraints on lensing quantities beyond a given family of mass models.
This work does not state whether or not certain assumptions on the global deflector mass distribution are valid for specific science investigations and their stated uncertainties.
One globally imposed constraint that is valid for any physical deflector mass distribution is the continuity in the deflection field. A set of local arc models do not demand this continuity between the different arcs, as, for example, illustrated for the example presented in Section~\ref{sec:example_mock} in Figure~\ref{fig:example_reconstruction} by the disconnected critical curves.
When the individual curved arcs are sufficiently separated from each other without constraining data in between, dropping the continuity assumption is a practical convenience for being agnostic to the deflection field behavior outside the data constraining region, and counting on a physical model that is able to continuously connect the different local regions.

In many real lensing configurations, arcs extend over large azimuthal angles, effectively adding more constraints on the azimuthal structure of the deflector and physically demanding stronger assumptions on the continuity of the global deflection field. More extended arcs most likely require also more and higher order local differentials to match the observations. In particular, differentials associated with the curvature strength and direction, which have not been considered in this work, may be required. Extensions to higher orders can be implemented within the provided framework and do not impact the general methodological questions and conclusions presented in this work. Continuity constraints and priors can also be added, either directly in the inference of curved arc constraints on the data, or in post-processing on the posterior level. Continuity can be demanded in both, the lensing differentials, as well as the total deflection. The latter is effectively demanding the positional constraints on the arcs to be a solution of a global macro model for a single source position.

In the case of fully connected arcs, effectively Einstein rings, require full curl-free continuity in the local deflection field in azimuthal direction. In this regime, where a subset of the source is displayed along a continuous rotation of the tangential direction, the shape-noise degeneracy is most effectively been broken.

However, we stress that invariances under the MST remain even in the regime of fully connected arcs, particularly impacting the constraining power in the radial direction.

\section{Conclusion} \label{sec:conclusion}

In this work, we introduced a formalism to describe the gravitational lensing distortion effects of curved extended arcs based on the eigenvectors and eigenvalues of the local lensing Jacobian and their directional differentials.
We identified a set of non-linear extended deflector descriptions that inherit the local properties able to describe the extent of individual lensed images.
Our parameterization is tightly linked to observable features in extended sources and allows one for an accurate extraction of the relevant information of extended images without imposing an explicit global deflector model.

We re-formulate the most general lensing invariance in an operator notation and subsequently quantify what aspects can be broken based on specific assumptions on the local lensing nature and assumed intrinsic source shape.

Our main findings are:
\begin{enumerate}
	\item The non-linear lensing nature in curved arcs allows one to partially break the shape-noise degeneracy. In particular, shape noise off-axis to the eigenvector directions can be constrained while on-axis shape distortions are more degenerate with lensing eigenvalues and curvature.
	\item Elliptical mass distributions lead to tangential stretch gradients but keep the curvature radius along the azimuth constant. External shear distortions do, in addition to  tangential stretch gradients, lead to offsets in the curvature radius and direction along the azimuth (Section~\ref{sec:tang_constraints}).
	\item Information on the radial mass profile can be obtained by measuring the differential thickness of arcs along the radial direction, the radial distance ratio of image pairs, or, when imposing azimuthal constraints, by the tangential stretch change along the radial direction (Section~\ref{sec:rad_constraints}).
	\item Imposing symmetries on the global form of the tangential behavior of the deflector profile can break the shape noise while imposing functional forms on the radial deflector profile can break the MST.
\end{enumerate}

Our formalism is applicable in all regimes of gravitational lensing, from the weak linear regime, the semi-linear regime up to the highly non-linear regime of highly magnified arcs and Einstein rings of multiple images.
The methodology presented in this work provides a framework to assess systematics, and to guide an inference effort in complexity choices based on the data at hand. Implementations of all the aspects presented in this work are available in \textsc{lenstronomy}.
% baseline for future exploration
The specific examples and discussions provided in this work can serve as a baseline for more extended theoretical and practical investigations and assessments in different regimes of gravitational lensing and for different scientific investigations. We outline applications and implications for dark matter substructure inferences, measuring the Hubble constant, and large scale structure inferences from the statistics of gravitational lenses.

\section*{Acknowledgments}
SB thanks Anowar Shajib, Dominique Sluse, Daniel Gilman, Tommaso Treu, Martin Millon, Lyne van de Vyvere and Roger Blandford for useful feedback in the process of writing this manuscript. Support for this work was provided by the National Science Foundation through NSF AST-1716527.

%%%%%%%%%%%%%%%%%%%%%%%%%%%%%%%%%%%%%%%%%%%%%%%%%%



%%%%%%%%%%%%%%%%%%%%%%%%%%%%%%%%%%%%%%%%%%%%%%%%%%

%%%%%%%%%%%%%%%%% APPENDICES %%%%%%%%%%%%%%%%%%%%%

\appendix

\section{Power-law elliptical mass distribution (PEMD)}\label{app:pemd}

\subsection{Parameterization}\label{app:pemd_parameterization}

The elliptical power-law mass distribution can be defined as\footnote{This is the current \textsc{lenstronomy} convention with version 1.8.1 and previous versions.}
\begin{equation}\label{eqn:epl_q}
\kappa(\theta_1, \theta_2) = \frac{3 - \gamma'}{2} \left(\frac{\theta_{\rm E}}{\sqrt{q \theta_1^2 + \theta_2^2/q}} \right)^{\gamma' - 1},
\end{equation}
where $q$ is the semi-minor to semi-major axis ratio, $\theta_{\rm E}$ is the Einstein radius, and $\gamma'$ is the logarithmic slope of the three-dimensional mass profile.
$\gamma'=2$ is an isothermal profile, the limit of $\gamma' \rightarrow 3$ results in a point mass and $\gamma' \rightarrow 1$ describes a uniform critical mass sheet.
The coordinates $(\theta_1, \theta_2)$ are rotates such that $\theta_1$ is along the semi-major axis.

Alternatively, the same profile can be defined as
\begin{equation}\label{eqn:epl_e}
\kappa(\boldsymbol{\theta}) = \frac{3 - \gamma'}{2} \left(\frac{\theta'_{\rm E}}{\theta_{r} \sqrt{1 - \epsilon\cos (2\phi)}} \right)^{\gamma'- 1},
\end{equation}
with $\theta_r$ is the radial distance to the center, $\phi$ is the angle relative to the major axis, and $\epsilon$ is the ellipticity, which is related to the axis ratio, $q$, by
\begin{equation}
	\epsilon = \frac{1 - q^2}{1 + q^2}.
\end{equation}
To provide an identical normalization of the deflection angles, the Einstein radii of expression (\ref{eqn:epl_q}), $\theta_{\rm E}$, and of expression (\ref{eqn:epl_e}), $\theta'_{\rm E}$,need to follow the relation
\begin{equation}
	\left(\frac{\theta'_{\rm E}}{\theta_{\rm E}}\right)^{2} = \frac{2q}{1+q^2}.
\end{equation}
The Einstein radius definition of expression (\ref{eqn:epl_q}) is such that the square average of the deflection angle along the semi-major and semi-minor axis corresponds to $\theta_{\rm E}$, while expression (\ref{eqn:epl_e}) matches the Einstein radius $\theta'_{\rm E}$ in directions half-way between the semi-major and semi-minor axes.

Computations for deflection angles and lensing potential are provided by \cite{Barkana:1998, Tessore:2015}\footnote{Both computational methods to compute lensing properties are implemented in \textsc{lenstronomy}.}.

\subsection{Eigenvectors in spherical case}
In the spherical case of the PEMD profile (Eqn.~\ref{eqn:epl_q},~\ref{eqn:epl_e}), the deflection deflection angle and differentials are simple analytical expressions. The deflection angle in radial direction is given by
\begin{equation}
	\alpha(r) = \theta_{\rm E} \left(\frac{\theta_{\rm E}}{r}\right)^{\gamma'-2}
\end{equation}
with $\theta_{\rm E}$ is the Einstein radius and $\gamma'$ is the three-dimensional power-law slope of the mass profile. 

The tangential and radial eigenvalues are given by
\begin{equation}
	\frac{1}{\lambda_{\text{tan}}} = 1 - \left(\frac{\theta_{\rm E}}{r}\right)^{\gamma' -1}
\end{equation}
and
\begin{equation}
	\frac{1}{\lambda_{\text{rad}}} = 1 + \left(\gamma' -2 \right)\left(\frac{\theta_{\rm E}}{r}\right)^{\gamma' -1}.
\end{equation}

The radial differential of the tangential eigenvalue, $\partial_r\lambda_{\rm tan}$, is given by
\begin{equation}\label{eqn:dlambda_tan_drad_round}
	\partial_r\lambda_{\rm tan} = \frac{(1 - \gamma') \left(\frac{\theta_{\rm E}}{r}\right)^{\gamma'}}{\theta_{\rm E} \left(1 - \left(\frac{\theta_{\rm E}}{r}\right)^{\gamma' - 1}\right)^2},
\end{equation}
and the radial differential of the radial eigenvalue, $\partial_r\lambda_{\rm rad}$, is given by
\begin{equation}\label{eqn:dlambda_rad_drad_round}
	\partial_r\lambda_{\rm rad} = \frac{(1 - \gamma') (2 - \gamma') \left(\frac{\theta_{\rm E}}{r}\right)^{\gamma}}{\theta_{\rm E} \left(1 + \left(\frac{\theta_{\rm E}}{r}\right)^{\gamma' - 1} (\gamma'-2) \right)^2}.
\end{equation}

At the Einstein radius $\theta_{\rm E}$, we can express the MST invariant quantity $\partial_r \lambda_{\rm rad} /\lambda_{\rm rad}$ as
\begin{equation} \label{eqn:pl_constraints}
	\frac{\partial_r \lambda_{\rm rad}(\theta_{\rm E})}{\lambda_{\rm rad}(\theta_{\rm E})} = \frac{\gamma'-2}{\theta_{\rm E}},
\end{equation}
and the overall lensing scale invariant quantity $\xi_{\rm rad}$ (Eqn.~\ref{eqn:xi_inv}) is given by
\begin{equation} \label{eqn:pl_constraints_reduced}
	\xi_{\rm rad} = \gamma'-2.
\end{equation}
This relation reflects the fact that the MST-invariant observational constraint captured by $\xi_{\rm rad}$, when interpreted as a constant power-law mass density, constrain the power-law slope and effectively breaks the MST.

\subsection{Eigenvectors in elliptical case}
In the elliptical case, the eigenvectors and directions remain the same as for the round case, substituting the $r$ by the circularized expression $r'$ (i.e. denominator of expression~\ref{eqn:epl_q} or \ref{eqn:epl_e}).

In addition, a non-zero second order differential emerges, namely the tangential stretch differential in tangential direction, $\partial_t\lambda_{\rm tan}$.
Using the chain rule, we can write
\begin{equation}\label{eqn:dtan_dtan_chain_rule}
	\partial_t\lambda_{\rm tan} = \frac{\partial \lambda_{\rm tan}}{\partial r'}\frac{\partial r'}{\partial e_{\rm tan}}.
\end{equation}
Adopting the ellipticity definition of the form of expression~(\ref{eqn:epl_e}), such that
\begin{equation}
	r' = r \sqrt{1 -\epsilon \cos(2\phi)},
\end{equation}
we can write
\begin{equation} \label{eqn:direction_diff}
	\frac{\partial r'}{\partial e_{\rm tan}} = \frac{\partial r'}{\partial \phi}\frac{\partial \phi}{\partial e_{\rm tan}} =  \frac{\epsilon \sin(2\phi)}{\sqrt{1 -\epsilon \cos(2\phi)}}.
\end{equation}
Combining Equations (\ref{eqn:dtan_dtan_chain_rule}) and (\ref{eqn:direction_diff}), we can compactly write
\begin{equation}\label{eqn:epl_dtan_dtan}
	\partial_t\lambda_{\rm tan} = \frac{\partial \lambda_{\rm tan}}{\partial r'} \frac{\epsilon \sin(2\phi)}{\sqrt{1 -\epsilon \cos(2\phi)}},
\end{equation}
with the first term given by the round case of expression~(\ref{eqn:dlambda_tan_drad_round}).

\subsection{Curved arc description with tangential stretch differential} \label{app:sie_arc}

The constant radial and tangential eigenvalue deflector model with constant tangential curvature is presented in Section~\ref{sec:curved_arc_basis} and can be written as a combination of a SIS and an MST. A convenient way to introduce a model satisfying the same quantities locally and has an additional tangential differential component $\partial_t\lambda_{\rm tan}$ is the singular isothermal ellipsoid (SIE) replacing the SIS profile. The SIE is the special case of the PEMD for $\gamma'=2$ (Appendix~\ref{app:pemd}) and $\partial_t\lambda_{\rm tan}$ is given by expression~\ref{eqn:epl_dtan_dtan}. To satisfy locally constraints on $\partial_t\lambda_{\rm tan}$ with the next leading order is set to zero is satisfied with an eccentricity off-axis by $\pi/4$ to the curvature direction. The off-axis direction is determined by the sign of imposed $\partial_t\lambda_{\rm tan}$.

\section{Ellipticity and shear} \label{app:ellipticity_shear}

\subsection{Shear}\label{app:shear}

Shear distortions are fully characterized by the constant $\gamma_1$ and $\gamma_2$ values (Eqn.~\ref{eqn:gamma1}, \ref{eqn:gamma2}) and lead to deflection filed
\begin{equation}
 \boldsymbol{\alpha}(\boldsymbol{\theta} ) = \left[\begin{array}{ c c } \gamma_1 & \gamma_2 \\ \gamma_2 & -\gamma_1 \end{array}\right] 
 (\boldsymbol{\theta - \theta_0}),
\end{equation}
where $\boldsymbol{\theta_0}$ is a, somewhat arbitrary, zero point of the deflection field, only impacting an overall constant shift of the deflection angle. 
In polar coordinates, we can also equivalently parameterize the shear distortions with an absolute shear strength $\gamma$ and an orientation relative to the first axis $\phi_{\gamma}$. The Cartesian shear components are then given by
\begin{align}
	\gamma_1 = \gamma \cos(2\phi_{\gamma}) &&
	\gamma_2 = \gamma \sin(2\phi_{\gamma}).
\end{align}

A pure shear distortions do have a magnification effect $\mu = (1-\gamma_1^2 -\gamma_2^2)^{-1}$ and thus do alter the size of the object in addition of causing distortions. Shape distortions agnostic to the intrinsic source size are generally referred to as reduced shear, and given by 
\begin{equation}
\gamma'_{1,2} = \frac{\gamma_{1,2}}{1 - \kappa}.
\end{equation}
In terms of a descriptive lens model, we require the knowledge of kappa in this notion. However, we can introduce a reduced shear model which, by design, has magnification $\mu=1$. Such a model, when defined by reduced shear components ($\gamma'_1$, $\gamma'_2$), requires a convergence of
\begin{equation}\label{eqn:kappa_reduced}
	\kappa = 1 - \frac{1}{\sqrt{1-\gamma_1^{'2} - \gamma_2^{'2}}}.
\end{equation}
We define the normalized reduced shear model, $\mathbf{L}_{\rm nrs}$, as the linear distortion model with parameters $\gamma'_1$ and $\gamma'_2$ where the convergence term is set by Equation~(\ref{eqn:kappa_reduced}). This specific linear distortion parameterization preserves the total magnification. The inverse lensing operator is given by the same operator with flipped signs in the reduced shear components
\begin{equation}
	\mathbf{L^{-1}_{\rm nrs}}(\gamma'_1, \gamma'_2) = \mathbf{L_{\rm nrs}}(-\gamma'_1, -\gamma'_2).
\end{equation}
This specific from of the shear description becomes relevant in Appendix~\ref{app:ellipticity} when discussing intrinsic surface brightness ellipticity and degeneracies with shear distortions.

\subsection{Ellipticity}\label{app:ellipticity}
A convenient way to describe elliptical surface brightness distributions is by the axis ratio $q$ and orientation $\phi_I$ of annuli of constant surface brightness.
A surface brightness profile with constant ellipticity can be described as a distortion transform of a radial surface brightness profile $I_r(r)$, with ellipticity operator $\mathbf{E}(x, y)$, such that $I_e = I_r(E(x, y))$.
Different ellipticity operators are used in the literature. Differences exist in the definition of the ellipticity as well as the overall size change. We are using the operator
\begin{equation} \label{eqn:ell_op}
	\mathbf{E}(q): (x, y) \rightarrow \left(\sqrt{q} x, y/\sqrt{q}\right)
\end{equation}
where $x$ is in the orientation of the major axis. This is the same ellipticity operator as used in the PEMD profile defined by expression~(\ref{eqn:epl_q}). The operator form of expression~(\ref{eqn:ell_op}) conserves the product-averaged radius.

A convenient basis to express the axis ratio $q$ and the orientation angle $\phi_I$ is with the eccentricity moduli
\begin{align}\label{eqn:eccentricity_definition}
	e_1 = \frac{1 - q}{1 + q} \cos(2\phi_I) &&
	e_2 = \frac{1 - q}{1 + q} \sin(2\phi_I).
\end{align}

\subsection{Shape noise} \label{app:shape_noise}

In the basis of the eccentricity moduli (Eqn.~\ref{eqn:eccentricity_definition}), the ellipticity operator $\mathbf{E}$ is identical to the lensing distortion operator $\mathbf{L_{\rm nrs}}$ (see Section~\ref{app:shear}) by identifying $\gamma_1 = e_1$ and $\gamma_2 = e_2$.

Using these bases for shear and ellipticity, we can identify the shape noise component of the SPT as
\begin{equation}
	\mathbb{1} = \mathbf{L_{\rm nrs}}(-e_1, -e_2) \mathbf{\circ} \mathbf{E}(e_1, e_2),
\end{equation}
and enables a separability of the MST component and shape noise component.

% elliptical transform of light profiles

% Don't change these lines
%\bsp    % typesetting comment
%\label{lastpage}

\bibliography{BibdeskLib}{}
\bibliographystyle{aasjournal}



\end{document}